\documentclass[prl,floatfix,twocolumn,superscriptaddress,showpacs,preprintnumbers,longbibliography,nofootinbib]{revtex4-2}
\usepackage{color}
\usepackage[utf8]{inputenc}
\usepackage[usenames,dvipsnames]{xcolor}
\usepackage{amsmath}
\usepackage{amssymb}
\usepackage{graphicx}
\graphicspath{{graphics/}}
\usepackage{epsfig}
\usepackage{dcolumn}
\usepackage{bm}
\usepackage{mathrsfs}
\usepackage{multirow}
\usepackage[all]{xy}
\usepackage{pbox}
\usepackage{lipsum}
\usepackage{nicefrac}
\usepackage{verbatim}
\usepackage{braket}
\usepackage{dsfont}
\usepackage{array}
\usepackage{makecell}
\usepackage{tabularx}
\usepackage{isotope}
\usepackage{xr}
\usepackage{amsthm} 
\usepackage[normalem]{ulem} 
\usepackage{physics}
\usepackage{float}
\usepackage{placeins}
\usepackage{svg}

\usepackage[colorlinks=true,linkcolor=blue,urlcolor=blue,citecolor=blue,pdfusetitle]{hyperref}

\usepackage{tikz}

\begin{document}
\title{Training the classification capability of large-scale quantum cellular automata}
\author{Mario Boneberg}
\affiliation{Institut f\"ur Theoretische Physik, Universit\"at Tübingen and Center for Integrated Quantum Science and Technology, Auf der Morgenstelle 14, 72076 T\"ubingen, Germany}
\author{Simon Kochsiek}
\affiliation{Institut f\"ur Theoretische Physik, Universit\"at Tübingen and Center for Integrated Quantum Science and Technology, Auf der Morgenstelle 14, 72076 T\"ubingen, Germany}
\author{Gabriele Perfetto}
\affiliation{Institut f\"ur Theoretische Physik, Universit\"at Tübingen and Center for Integrated Quantum Science and Technology, Auf der Morgenstelle 14, 72076 T\"ubingen, Germany}
\author{Igor Lesanovsky}
\affiliation{Institut f\"ur Theoretische Physik, Universit\"at Tübingen and Center for Integrated Quantum Science and Technology, Auf der Morgenstelle 14, 72076 T\"ubingen, Germany}
\affiliation{School of Physics and Astronomy and Centre for the Mathematics and Theoretical Physics of Quantum Non-Equilibrium Systems, The University of Nottingham, Nottingham, NG7 2RD, United Kingdom}


\begin{abstract}
In the vicinity of a phase transition ergodicity can be broken. Here, different initial many-body configurations evolve towards one of several fixed points, which are macroscopically distinguishable through an order parameter. This mechanism enables state classification in quantum cellular automata and feed-forward quantum neural networks. 
We demonstrate that this capability can be efficiently learned from training data even in extremely high-dimensional state spaces. We illustrate this using a quantum cellular automaton that allows binary classification, which is closely connected to the dynamics of a $\mathbb{Z}_2$-symmetric Ising model with local interactions and dissipation. This approach can be generalized beyond binary classification and offers a natural framework for exploring the link between emergent many-body phenomena and the interpretation of data processing capabilities in the context of quantum machine learning.
\end{abstract}
\maketitle
\noindent {\bf Introduction.---}
\begin{figure}[t]
    \centering
    \includegraphics[width=\linewidth]{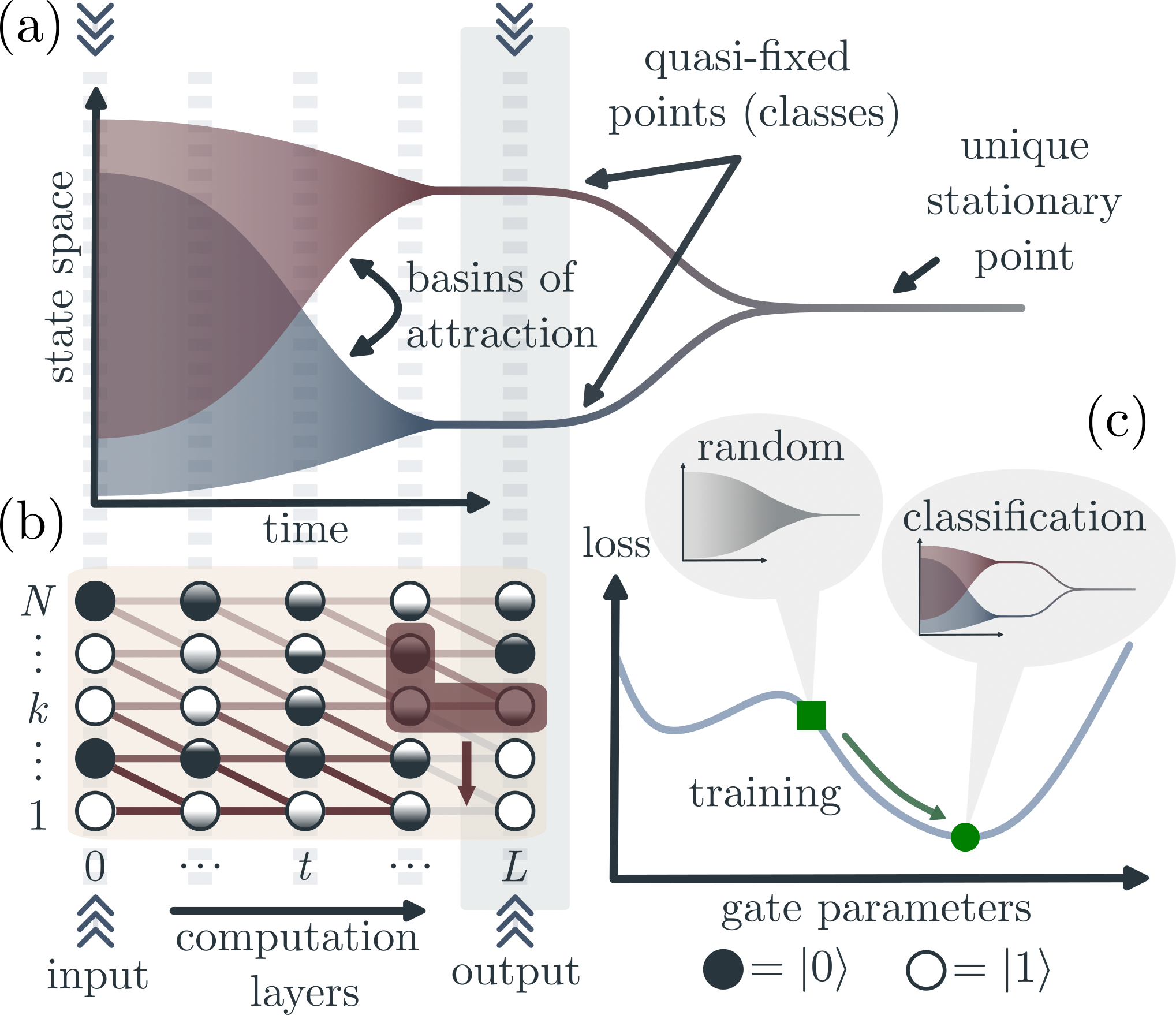}
    \caption{\textbf{Transient ergodicity breaking, quantum cellular automaton, classification and training.} (a) Transient non-ergodic dynamics evolves clusters of initial states (basins of attraction) towards quasi-fixed points (here two) at intermediate times. For finite system size, in the long-time limit, all states converge to a unique stationary fixed point and ergodicity is restored. (b) QCA are defined on a $2d$ lattice of qubits. Given an input state on the first layer, quantum gates (red shape and lines) are applied sequentially along the layer. This is repeated layer-by-layer from left to right and produces an output state on the last layer. The horizontal axis can be interpreted as a time-axis and 
    each layer can be interpreted as a discrete-time instance of a system evolving under a dissipative quantum many-body dynamics. Choosing the number of layers such that the output state matches with the non-ergodic window (grey rectangle), realizes a binary classification of inputs. Note, that shaded circles are used to sketch qubits in quantum states such as, e.g., superpositions. (c) Starting from a random initialization of the gates (green square), training the QCA yields gate parameters that enable classification of inputs (green circle).}
    \label{fig:1}
\end{figure}
In many-body systems a variety of interesting physical phenomena manifest as a consequence of the interactions between a large number of elementary constituents \cite{bardeen1957,anderson1958,hopfield1982}. Paradigmatic in this context are phase transitions \cite{hinrichsen2000,landau2013,sachdev2016,huang2008}, where order is abruptly established when tuning an external parameter across a critical value. This is usually accompanied by spontaneous symmetry breaking and nonergodicity \cite{strocchi2005,huang2008,itzykson1989,ruelle1969,baxter2016}, where initial states, that are located in different basins of attraction [see Fig.~\ref{fig:1}(a)], dynamically relax towards different fixed points. A well-known example is the Ising model \cite{ising1925,onsager1944,baxter2016}, which in its ferromagnetic phase can relax to two different states with opposite magnetization. In finite systems, such a strict ergodicity breaking is fundamentally forbidden \cite{itzykson1989,ruelle1969,goldenfeld2018,davies1982,mccraw1978}. Nevertheless, ergodicity can be broken during transient periods \cite{mori2018,yin2025,macieszczak2016,macieszczak2021,mezard1987,davies1982,mccraw1978} where the dynamics may feature multiple \textit{quasi-fixed points} [see Fig.~\ref{fig:1}(a)]. 

The phenomenon of ergodicity breaking is closely related to data processing and classification. 
A prominent example is the classical Hopfield neural network \cite{hopfield1982,amit1985a,amit1985b,amit1987} whose (quasi-)fixed points are interpreted as patterns or associative memories, which are dynamically retrieved. Moreover, in quantum machine learning \cite{biamonte2017,schuld2021,schuld2022,cerezo2022}, certain classes of gate-based quantum cellular automata (QCA) \cite{farrelly2020,arrighi2019,lesanovsky2019,gillman2020,gillman2021a,gillman2021b,nigmatullin2021,gillman2022a,guedes2024,wagner2024,wagner2025} [see Fig.~\ref{fig:1}(b)] and quantum neural networks \cite{beer2020,bondarenko2020,lewenstein2021,beer2022,gillman2022b,boneberg2023,locher2023,sutter2025,boneberg2025} process input states via an effectively 
dissipative quantum-many body evolution \cite{breuer2002,rivas2012}. This can feature phase transitions and breaking of microscopic weak symmetries in the thermodynamic limit \cite{bartolo2016,minganti2023,lieu2020,buca2012,albert2014,sieberer2025}. So far, however, it is not fully understood how breaking of microscopic symmetries can possibly manifest in QCA and how it can be beneficially exploited to train the capability of classifying data.   


In this paper, we  answer this question by developing an overarching connection between weak symmetries of dissipative quantum dynamics, transient ergodicity breaking and the trainability of the data classification on a QCA. We illustrate our ideas, using a QCA whose architecture is inspired by a dissipative quantum Ising model with short-ranged interactions and a microscopic weak $\mathbb{Z}_2$-symmetry \cite{buca2012,albert2014,overbeck2017}. It features, in the thermodynamic limit, a continuous transition from an ergodic to a $\mathbb{Z}_2$ symmetry-broken phase. In the QCA --- which is necessarily of finite size --- we show that this phenomenon manifests in a transient ergodicity breaking through the emergence of two quasi-fixed points. These can be exploited for binary classification [cf. Fig.~\ref{fig:1}(a-b)]. We, moreover, demonstrate that the QCA can acquire this binary classification capability through training [see Fig.~\ref{fig:1}(c)]. Our physics-informed QCA benefits from local gates and a loss function that has the form of an order parameter. This architecture appears to prevent trainability problems typically arising in general quantum machine learning architectures, e.g., barren plateaus \cite{mcclean2018,cerezo2021b,sharma2022,cerezo2023,larocca2024,ragone2024}. Using simulations based on tensor-networks  \cite{orus2014,paeckel2019,gillman2021b,boneberg2025}, we demonstrate that these ingredients yield a structured optimization landscape for the loss function even for a Hilbert space dimension of up to $10^{24}$. Our findings may thus provide insights into how to design trainable architectures for large-scale quantum machine learning. 

\noindent {\bf Physics-informed quantum cellular automaton.---}
We consider a QCA \cite{lesanovsky2019,gillman2020,gillman2021a,gillman2021b,gillman2022a,gillman2022b,boneberg2023,boneberg2025} defined on a two-dimensional lattice with $N$ vertical and $L+1$ horizontal sites [cf. Fig.~\ref{fig:1}(b)]. With each site we associate a qubit with basis states denoted by $\ket{0}$ (vacuum) and $\ket{1}$. On the lattice, a vertical slice, indexed by the time coordinate $t$, is referred to as layer of the QCA. The first layer ($t=0$) is prepared in some input state $\rho_0$, while all the other layers are initialized in the vacuum state. Quantum gates $G_k$, acting on two adjacent layers in a neighborhood of the $k$-th vertical sites, are applied sequentially along the layer and together form the two-layer global gate $\mathcal{G}= \prod_{k=1}^N G_k$ (up to boundary terms). This global gate is then applied successively along the horizontal direction, layer-by-layer, and propagates the initial state $\rho_0$ to a final state $\rho_L$ at time $t=L$ on the last layer. The horizontal axis can be interpreted as a time-axis and the above construction defines a (discrete-time) dynamics on the lattice via the recurrence relation \cite{gillman2022b,boneberg2023}:
\begin{equation} \label{recurrence}
    \rho_t = \Tr_{t-1}\left(\mathcal{G} \rho_{t-1} \otimes \outerproduct{\mathbf{0}}{\mathbf{0}}   \mathcal{G}^\dagger \right).
\end{equation}
This relation evolves a reduced state $\rho_{t-1}$ from one time layer, $t-1$, to the next, $t$. Here, the trace is taken over the full layer $t-1$ and the operator $\outerproduct{\mathbf{0}}{\mathbf{0}}$ represents the vacuum state on layer $t$. 

To link the dynamics of QCA with the evolution of a dissipative many-body quantum system, we choose the translation-invariant and time-independent local gate \cite{boneberg2023,boneberg2025} as
\begin{equation} \label{localgate}
    G_{k} = \mathrm{SWAP}_{k} \times e^{-i \sqrt{\delta t} \big( J_{k} \otimes \sigma_{k}^+ + \text{h.c.}\big)} \times e^{-i \delta t H_{k} \otimes \mathds{1}}.
\end{equation}
Here, the operator $\sigma^+_k = \sigma^x_k + i \sigma^y_k$  is defined via the Pauli spin matrices $\sigma^\alpha_k, \alpha =x,y,z$, where the index $k$ refers to the lattice site $k$.
The rightmost term generates a coherent evolution on layer $t-1$ under the local \emph{Hamiltonian} $H_k$ with time-step $\delta t$. The middle term includes the \emph{jump operators} $J_k$. This also acts non-trivially on layer $t$ and induces dissipation after performing the trace. Finally, the $\mathrm{SWAP}_k$ operator interchanges the state of the qubits located on sites $(k,t-1)$ and $(k,t)$.

With this choice of the gates, Eq. \eqref{recurrence} evolves a the state of one layer to the next according to  $\rho_t\approx \exp (\mathcal{L}\delta t)[\rho_{t-1}]$ for $\delta t \ll 1$\cite{boneberg2023,boneberg2025,lorenzo2017,ciccarello2017,ciccarello2022,cattaneo2021,cattaneo2022}. Hence each layer may be interpreted as an instance of a one-dimensional spin-$\nicefrac{1}{2}$ chain of length $N$ evolving under a \emph{Markovian} dissipative quantum dynamics generated by the so-called Lindblad map \cite{lindblad1976,gorini1976,breuer2002,rivas2012}
\begin{align}\label{lindbladian}
    \mathcal{L}[\bullet] =& -i\left[\sum_{k=1}^N H_k, \bullet \right] + \sum_{k=1}^N \left( J_{k} \bullet J_{k}^\dagger - \frac{1}{2} \left\{ J_{k}^{\dagger} J_{k}, \bullet \right\} \right).
\end{align}
This establishes the connection between the dynamics of a dissipative many-body quantum system and information processing within QCA and related quantum neural network architectures that are described by Eq. (\ref{recurrence}) \cite{beer2020,beer2022,gillman2022b, boneberg2023,boneberg2025}.

\noindent {\bf Phase transitions, ergodicity breaking and classification.---} This link now allows us to understand how weak symmetries, phase transitions and associated ergodicity breaking constitute the physical mechanism that lies at the heart of the data classification capabilities of QCA [cf. Fig.~\ref{fig:1}(a,b)]. To make this case, we focus here on a specific QCA architecture, where the gates \eqref{localgate} are inspired by a dissipative Ising model \cite{overbeck2017}. The local Hamiltonian and jump operators are
\begin{equation}\label{ising}
    H_k = \frac{\Omega}{2} \sigma_k^z - \frac{V}{4} \sigma_k^x \sigma_{k+1}^x \ , \qquad J_k= \sqrt{\kappa} \sigma_k^- ,
\end{equation}
where $\Omega$ is a transverse field and $V$ parametrizes the  nearest-neighbor interaction between the spins. The jump operators model local decay in the $z$-basis at a rate $\kappa$.
The Lindblad map \eqref{lindbladian} and \eqref{ising} has a so-called \emph{weak} $\mathbb{Z}_2$ symmetry \cite{buca2012,albert2014,minganti2018}, meaning that the super-operator $\mathcal{P}[\bullet] = \prod_{k=1}^N \sigma_k^z \bullet \prod_{k=1}^N \sigma_k^z$ commutes with the Lindblad generator, $[\mathcal{L}, \mathcal{P}]=0$. This symmetry will later underpin the data classification capability of the QCA.

To illuminate the general mechanism, it is in fact convenient to work with the Lindbladian \eqref{lindbladian} rather than the discrete evolution under Eq. (\ref{recurrence}). First, we perform a \emph{mean-field} approximation by restricting ourselves to the manifold of factorized and translation-invariant states 
$\rho^{\mathrm{MF}}_t = \prod_{k=1}^N \rho^{(k)}_t$. Second, we compute the time evolution of the magnetization components $\mu =x,y,z$:
\begin{equation} \label{orderparameter}
    m_t^\mu  = \Tr(\frac{1}{2N}\sum_{k=1}^N \sigma_k^\mu\,  \rho_t).
\end{equation}
They obey a closed set of mean-field equations of motion, which read (see Supplemental Material \cite{SM}):
\begin{subequations}
\begin{align}
    \Dot{m}_t^x &= - \Omega m_t^y - \frac{\kappa}{2} m_t^x,\\
    \Dot{m}_t^y &= \Omega m_t^x +2 V m_t^x m_t^z  -\frac{\kappa}{2} m_t^y, \\
    \Dot{m}_t^z &= -2 V m_t^x m_t^y  - \kappa \left( m_t^z + \frac{1}{2} \right) .
\end{align}
\label{eq:mean_field}
\end{subequations}
The stationary value of the $x$-component, $m_\infty^x$, is a suitable order parameter as it allows to distinguish dynamical fixed points. Further below, we exploit this property in our training protocol of the QCA to define a loss function, which can be efficiently evaluated [see Fig.~\ref{fig:1}(c) and \ref{fig:4}(c)].

\begin{figure}[t]
    \centering
    \includegraphics[width=\linewidth]{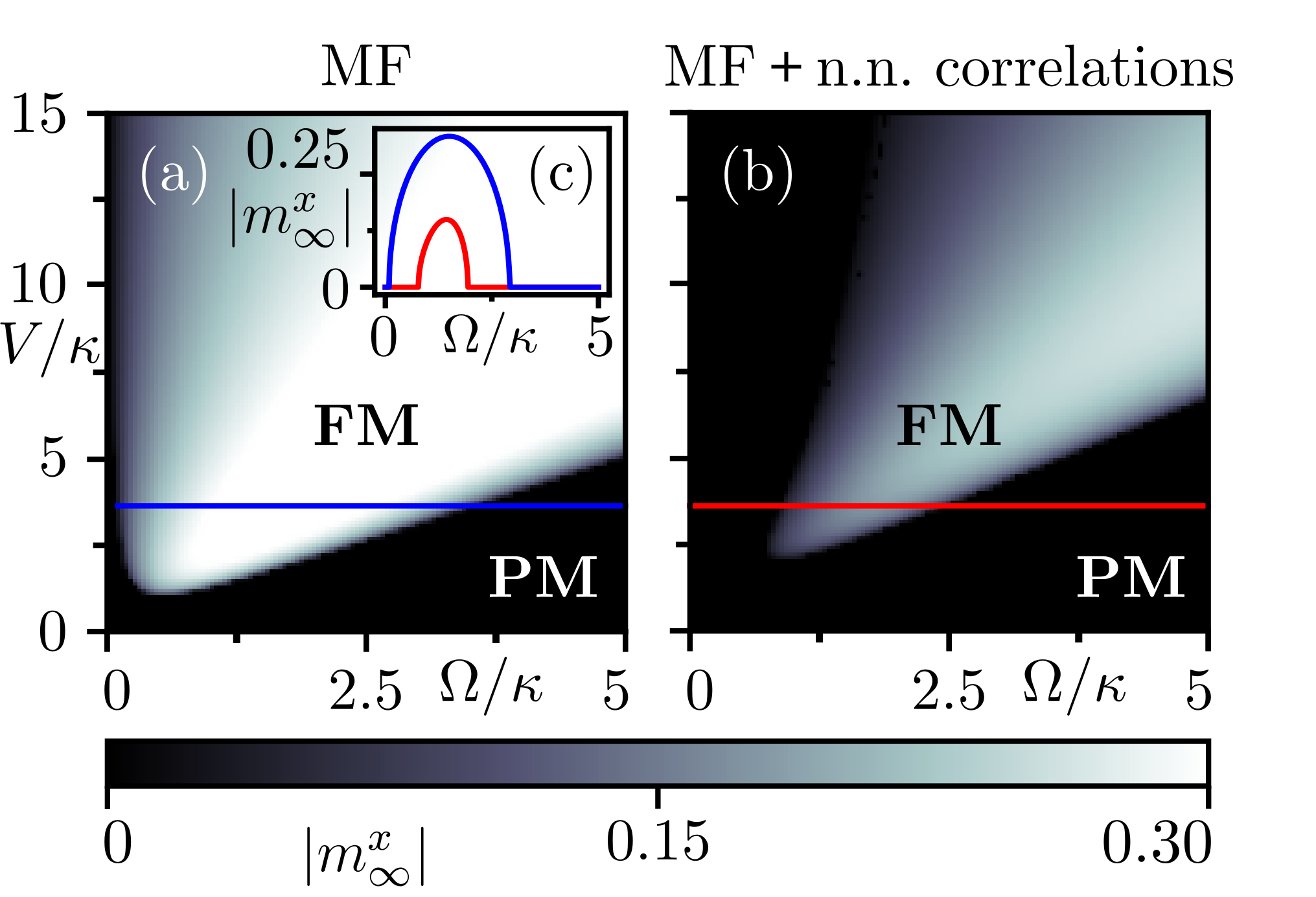}
    \caption{\textbf{Phase diagram in one dimension.} (a) Absolute value of the stationary order parameter ($x$-magnetization, $|m_{\infty}^x|$), as a function of dimensionless field strength $\Omega / \kappa$ and interaction strength $V / \kappa$ in one dimension. Numerically solving the mean-field (MF) Eqs.~\eqref{eq:mean_field}, we find a parameter regime, where $|m_{\infty}^x|$ acquires a finite value. In this ferromagnetic (FM) phase, the order parameter $m_{\infty}^x$ can either have positive or negative sign, i.e., the $\mathbb{Z}_2$ symmetry is broken. 
    (b) Same as panel (a), but for numerically solving the equation of motion coupling the order parameter with nearest-neighbor (n.n.) correlation functions (MF + n.n. correlations). Correlations reduce the extension of the FM phase. In panel (c), we plot a cut of the phase diagram at $V=3 \kappa$. The transition from the paramagnetic (PM) to the FM phase is continuous both in mean-field and when nearest neighbor correlations are included.}
    \label{fig:2}
\end{figure}

As illustrated in Fig.~\ref{fig:2}(a), the mean-field equations feature two qualitatively different regimes. 
For sufficiently large values of $\Omega/\kappa$ and $V/\kappa$ one finds a \textit{ferromagnetic} symmetry-broken phase, where the stationary order parameter $m_{\infty}^x$ assumes a nonzero value with either positive or negative sign. Which one of the two values is selected, depends on the initial condition and thus ergodicity is broken. Elsewhere in the phase diagram, one finds a unique \textit{paramagnetic} stationary solution where $m_{\infty}^x=0$. 
In systems with one spatial dimension, which applies to our QCA, strong fluctuations may lead to deviations from mean-field theory. To show that the phase transition is in fact robust to these, 
we include nearest-neighbor correlations.  
This procedure, which is discussed in Ref. \cite{PRA_BBGKY} and the Supplemental Material \cite{SM}, yields the phase diagram shown in Fig.~\ref{fig:2}(b). The inclusion of correlations shrinks the ferromagnetic phase, but does not alter the nature of the phase transition. This is manifest in Fig.~\ref{fig:2}(c), where we show cuts through the two phase diagrams.
In both cases the phase transition is continuous and originates from the breaking of the $\mathbb{Z}_2$-symmetry. 

\begin{figure}[t]
    \centering
    \includegraphics[width=\linewidth]{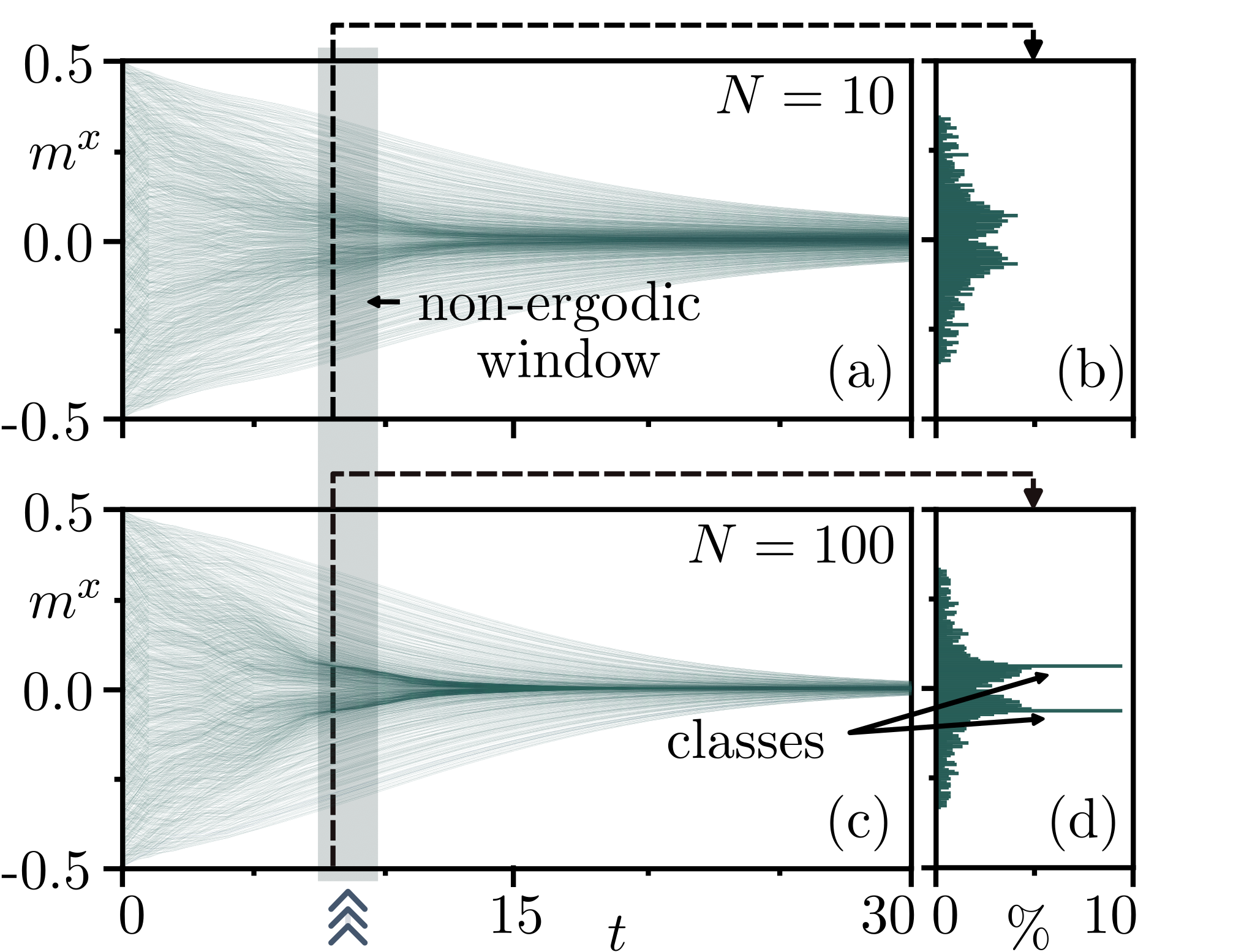} 
\caption{\textbf{Transient ergodicity breaking and bimodality in Ising model-informed quantum cellular automaton.} (a) Evolution of the order parameter $m_t^x$ through the QCA defined by Eqs. \eqref{localgate} and  \eqref{ising}, choosing $\Omega = 3 \kappa, V=15 \kappa$ and $N=10$. Shown are $2000$ magnetization curves, obtained from evaluating Eq.~\eqref{recurrence} using matrix product operators and states with bond dimensions $\chi_{\mathrm{MPO}}=16$ and $\chi_{\mathrm{MPS}}=48$, respectively \cite{SM}. The curves concentrate near two quasi-fixed points during a transient non-ergodic time window. (b) The histogram of the magnetizations at layer $t=8$ shows bimodality. (c) Same as (a), but for $N=100$. (d) The histogram of magnetizations at $t=8$ exhibits more pronounced bimodality with higher and sharper peaks. We have chosen a time-step size $ \kappa \delta t=0.1$ in these plots. The bins in the histograms have size $0.005$.}
    \label{fig:3}
\end{figure}

\begin{figure*}[t]
    \centering
    \includegraphics[width=\textwidth]{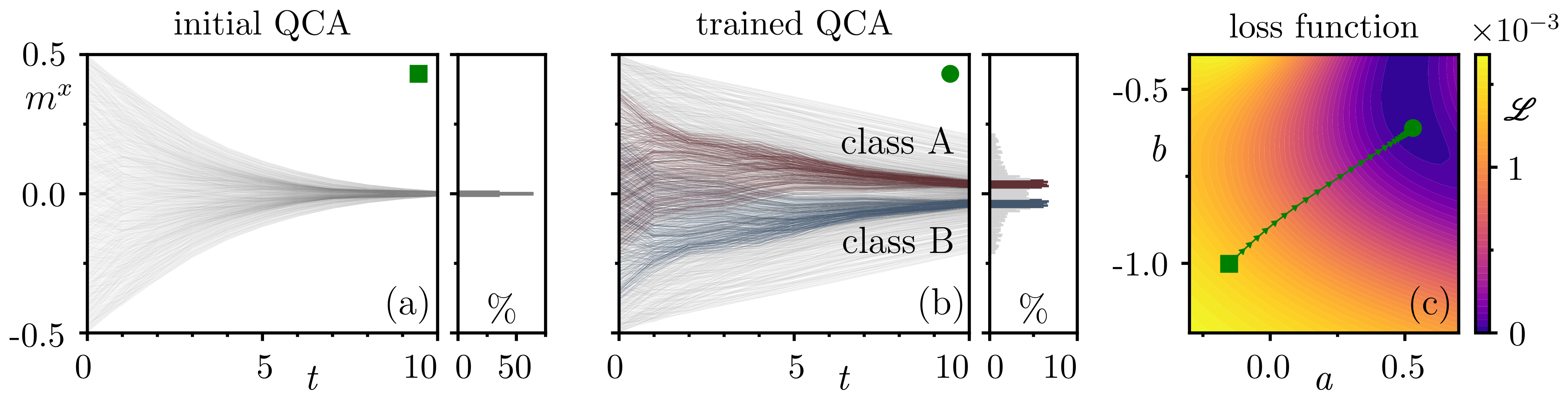}
    \caption{\textbf{Training classification capability of quantum cellular automaton.} (a) Evolution of the order parameter $m_t^x$ across the layers of the untrained QCA for $2000$ uniformly sampled initial states $\rho_0$ with magnetizations $m_0^x \in [-0.5,0.5]$ (left panel). The histogram of output magnetizations (right panel) $m_{10}^x$ displays a single peak. (b) Same as (a), but for the trained QCA. On the output layer the magnetization curves concentrate around two values, shown in red and blue, respectively, and the output histogram shows two clear peaks. The QCA thus realizes a binary classification. We trained for $100$ update repetitions with a learning rate $\epsilon = 80$. (c) Density plot of the order-parameter based loss \eqref{loss} as a function of the two training parameters $a, b$ for $N=20$. The loss exhibits a broad minimum which supports efficient training. Starting from the untrained QCA with large loss (green square) the training protocol enables learning of parameters (green circle) where the loss is minimal. The green line indicates the individual parameter updates during the training protocol. In our simulations we use a maximum matrix product operator and state bond dimensions of $\chi_{\mathrm{MPO}}=16$ and $\chi_{\mathrm{MPS}}=48$, respectively \cite{SM}. We have chosen a bin size $0.005$ for the histograms. }
    \label{fig:4}
\end{figure*}

The QCA defined by Eq. \eqref{recurrence} and gates \eqref{localgate} with Hamiltonian and jump operators (\ref{ising}) implements a finite-size (and discrete-time) version of the model we have just discussed.  
In such setting it is known that symmetries cannot be spontaneously broken \cite{itzykson1989,ruelle1969,goldenfeld2018,davies1982,mccraw1978} and non-ergodic behavior is not expected. Nonergodicity can at best be a transient effect, which in fact turns also out to be the case in our QCA. Here, the two opposite values for the order parameter $m_{\infty}^x$ in the ferromagnetic phase translate into two quasi-fixed points of the dynamics. This is apparent in Fig.~\ref{fig:3}, where we display the value of the order parameter $m_t^x$ within each layer $t$ of the QCA. The data is computed via a numerical simulation of Eq.~\eqref{recurrence} based on tensor networks \cite{orus2014,paeckel2019,gillman2021b,SM,boneberg2025}.   
Each of the displayed curves corresponds to one of $2000$ initial states with initial order parameter values sampled from the interval $m_0^x \in [-0.5,0.5]$ (see supplemental material \cite{SM}). 
The magnetization curves indeed concentrate around two values (reflecting the two ferromagnetic states) within a transient \emph{non-ergodic window} [cf. Fig.~\ref{fig:3}(a) and (b)]. These quasi-fixed points are an emergent phenomenon in the sense that they become more pronounced as we increase the number $N$ of qubits per layer [see Fig.~\ref{fig:3}(c) and (d)]. At long times all curves approach a unique state where the order parameter becomes zero, illustrating that ultimately the symmetry is restored and the dynamics is ergodic. However, by choosing the output layer, i.e., the final time slice, of the QCA such that it coincides with the non-ergodic window, one finds, indeed, that different initial states propagate towards macroscopically distiguishable order parameter values. One may therefore interpret the quasi-fixed points as classes and hence the QCA provides a binary classification mechanism, as sketched in Fig.~\ref{fig:1}(a,b). 

\noindent {\bf Training classification capability.---}
The two ferromagnetic quasi-fixed points, and hence the classification capability, appear only for certain parameters choices, broadly in line with the mean-field phase diagrams in Fig. \ref{fig:2}. In the following, we aim to understand whether and how a QCA can achieve this capability through training. From the physics perspective, which we adopt in this work, this means that we ask whether the coupling constants of the operators (\ref{ising}) can be learned through training. This is not obvious, as already the identification of well-behaved loss functions is challenging, when dealing with high-dimensional state spaces: the Hilbert spaces we are considering have a dimension of up to $4^N \times 4^N$, which is of the order $\sim 10^{24}$ for a QCA of width $N=20$.
%

Our training is based on the training data that consists of $P$ pairs $\{ \rho_{\mathrm{in}}^{(\nu)}, m_{x, \mathrm{out}}^{(\nu)} \}_{\nu=1,\dots,P}$ of input states and output magnetizations. We obtain them by evolving a QCA which features quasi-fixed points up to the non-ergodic window [see Fig.~\ref{fig:3}(a,c)]. We then initialize a QCA which does not classify inputs, i.e., does not feature a non-ergodic window. To this end we use in Eq. \eqref{localgate} the jump operators
\begin{equation}
J_k (a,b)= a \sigma_k^x +ib \sigma_k^y,
\label{eq:gate_parametrization}
\end{equation}
which depend on two real parameters $a$ and $b$. This parametrization also includes the jump operators of Eqs. \eqref{ising}. For the Hamiltonians $H_k$ we choose the same form as in Eqs.~\eqref{ising} and fix $\Omega=3 \kappa$ and $V=15 \kappa$. The output states $\rho^{(\nu)}_{\mathrm{out}}$ of this untrained QCA, which are obtained by propagating the states $\rho_{\mathrm{in}}^{(\nu)}$ from the training set, have output magnetization that are in general different from the desired output magnetizations $m_{x,\mathrm{out}}^{(\nu)}$. We quantify this difference through the \textit{loss function} 
\begin{equation} 
\label{loss}
    \mathscr{L} = \frac{1}{P} \sum_{\nu=1}^P \Tr(\bigg( \frac{1}{2N} \sum_{k=1}^N \sigma_k^x  - m_{x,\mathrm{out}}^{(\nu)} \bigg) \rho_{\mathrm{out}}^{(\nu)})^2.
\end{equation}
The goal of training is to find the QCA gates in Eq. \eqref{localgate} such that this loss is minimized.
This is generally done by \emph{gradient descent} \cite{beer2020}. 
Following the algorithm of Ref.~\cite{boneberg2025}, this amounts in our case to updating the jump operators in parameter-space via 
\begin{equation} \label{update}
    J_k(a,b) \longrightarrow J_k(a,b) + \epsilon J_k(\tilde{a},\tilde{b}) = J_k(a',b'),
\end{equation}
with $a'= a+ \epsilon \tilde{a}$ and $b'= b+ \epsilon \tilde{b}$. The parameters $\tilde{a},\tilde{b}$ are computed such that we move into the direction of steepest descent of $\mathscr{L}$ under the update [green arrows in Fig. \ref{fig:4}(c)]. The learning rate $\epsilon$ controls the step-size in parameter-space and we have to repeat this update multiple times in order to successively move towards a minimum of $\mathscr{L}$. We do not update the Hamiltonians $H_k$. Note that the computation of the update parameters $\tilde{a},\tilde{b}$ requires the evolution of many instances of the QCA \cite{boneberg2025}. This is computationally highly demanding even when employing tensor network methods given the large Hilbert space dimension.

To show that the training indeed works, we choose a QCA depth of $L=10$. We start by generating four training pairs by evolving a QCA with $a=\nicefrac{1}{2}, b=-\nicefrac{1}{2}$ in Eq.~\eqref{eq:gate_parametrization}, which implements the $\sigma^-$ jump operators of the Ising model, Eq. (\ref{ising}). We then initialize an untrained QCA with parameters $a=-0.15, b=-1$, which does not classify inputs as can be seen in Fig.~\ref{fig:4}(a). After completing the training the distribution of output magnetizations is bimodal, as can be seen in Fig. \ref{fig:4}(b). This establishes that the QCA acquired binary classification capability. The training is enabled by the fact that in spite of the large state space, the loss landscape, visualized in Fig.~\ref{fig:4}(c), exhibits clearly defined structure with broad minima.  This allows the gradient descent to rapidly find the minimum (green circle) when starting from the untrained QCA (green square).
Key for this benign behavior appears to be the form of the loss function (\ref{loss}), which is constructed using the order parameter and is thus also well-behaved in large systems. This is in contrast to loss functions that are computed from the overlap between a final state and the desired target state \cite{bisio2010,cincio2018,mcclean2018,cerezo2020,beer2020,xiao2022,wu2024,ewald2025}. Here, training is generally known to suffer from \textit{barren plateaus} which are a consequence of structure-less loss landscapes without a clearly identifiable gradient \cite{anderson1967,mcclean2018,cerezo2021a,cerezo2021b,sharma2022,cerezo2023,larocca2024,ragone2024}. 

\noindent {\bf Summary and outlook.---} Our work builds on a link between dissipative quantum many-body dynamics and information processing in QCA and related quantum neural network architectures. This perspective allowed us to establish a direct connection between the symmetries of a many-body system and the ability of an associated QCA to classify data. Moreover, we have shown that employing a loss functions inspired by physical order parameters enables efficient training in large state spaces.
The structure of the specific QCA studied here is inspired by a $\mathbb{Z}_2$-symmetric dissipative quantum Ising model, which yields a binary classifier. Multiclass classification is similarly achievable using QCA that are constructed from many-body systems with higher symmetries \cite{potts1952,solyom1981a,solyom1981b}. There are a manifold of possible directions for future work. First, it would be interesting to investigate how the transient nonergodicity in our QCA links to the closing of the spectral gap \cite{minganti2018,esposito2024,causer2025} and metastability \cite{macieszczak2016,macieszczak2021}. Second, studying training in an extended parameter space, e.g., by local Hamiltonians and jump operators that depend on $k$ and $t$ (spatial and temporal inhomogeneity), would be highly desirable. Given the involved system sizes, this is numerically rather complex on classical computers and would thus constitute a use case for quantum hardware. Here, it would be interesting to understand whether the noise on NISQ devices could be mitigated or even exploited, given that the QCA dynamics is effectively already dissipative.

\begin{acknowledgements}
\noindent \textbf{Acknowledgments.} 
We thank F. Carollo for discussions. The research leading to these results has received funding from the Deutsche Forschungsgemeinschaft (DFG, German Research Foundation) under Project No. 449905436, as well as through the Research Unit FOR 5413/1, Grant No. 465199066. This Project has also received funding from the European Union’s Horizon Europe research and innovation program under Grant Agreement No. 101046968 (BRISQ), and from EPSRC under Grant No. EP/V031201/1. We acknowledge the financial support by the German Federal Ministry of Research, Technology and Space (BMFTR) within the project "Neuronale Quantennetzwerke auf NISQ-Quantencomputern (NeuQuant)" under grant 13N17065. IL is a member of the Machine Learning Cluster of Excellence, funded by the Deutsche Forschungsgemeinschaft (DFG, German Research Foundation) under Germany’s Excellence Strategy—EXC Number 2064/1 - Project Number 390727645. The authors acknowledge support by the state of Baden-Württemberg through bwHPC and the German Research Foundation (DFG) through grant no INST 40/575-1 FUGG (JUSTUS 2 cluster).
\end{acknowledgements}

\bibliography{Reference}

\begin{thebibliography}{88}%
\makeatletter
\providecommand \@ifxundefined [1]{%
 \@ifx{#1\undefined}
}%
\providecommand \@ifnum [1]{%
 \ifnum #1\expandafter \@firstoftwo
 \else \expandafter \@secondoftwo
 \fi
}%
\providecommand \@ifx [1]{%
 \ifx #1\expandafter \@firstoftwo
 \else \expandafter \@secondoftwo
 \fi
}%
\providecommand \natexlab [1]{#1}%
\providecommand \enquote  [1]{``#1''}%
\providecommand \bibnamefont  [1]{#1}%
\providecommand \bibfnamefont [1]{#1}%
\providecommand \citenamefont [1]{#1}%
\providecommand \href@noop [0]{\@secondoftwo}%
\providecommand \href [0]{\begingroup \@sanitize@url \@href}%
\providecommand \@href[1]{\@@startlink{#1}\@@href}%
\providecommand \@@href[1]{\endgroup#1\@@endlink}%
\providecommand \@sanitize@url [0]{\catcode `\\12\catcode `\$12\catcode
  `\&12\catcode `\#12\catcode `\^12\catcode `\_12\catcode `\%12\relax}%
\providecommand \@@startlink[1]{}%
\providecommand \@@endlink[0]{}%
\providecommand \url  [0]{\begingroup\@sanitize@url \@url }%
\providecommand \@url [1]{\endgroup\@href {#1}{\urlprefix }}%
\providecommand \urlprefix  [0]{URL }%
\providecommand \Eprint [0]{\href }%
\providecommand \doibase [0]{https://doi.org/}%
\providecommand \selectlanguage [0]{\@gobble}%
\providecommand \bibinfo  [0]{\@secondoftwo}%
\providecommand \bibfield  [0]{\@secondoftwo}%
\providecommand \translation [1]{[#1]}%
\providecommand \BibitemOpen [0]{}%
\providecommand \bibitemStop [0]{}%
\providecommand \bibitemNoStop [0]{.\EOS\space}%
\providecommand \EOS [0]{\spacefactor3000\relax}%
\providecommand \BibitemShut  [1]{\csname bibitem#1\endcsname}%
\let\auto@bib@innerbib\@empty
\bibitem [{\citenamefont {Bardeen}\ \emph {et~al.}(1957)\citenamefont
  {Bardeen}, \citenamefont {Cooper},\ and\ \citenamefont
  {Schrieffer}}]{bardeen1957}%
  \BibitemOpen
  \bibfield  {author} {\bibinfo {author} {\bibfnamefont {J.}~\bibnamefont
  {Bardeen}}, \bibinfo {author} {\bibfnamefont {L.~N.}\ \bibnamefont
  {Cooper}},\ and\ \bibinfo {author} {\bibfnamefont {J.~R.}\ \bibnamefont
  {Schrieffer}},\ }\bibfield  {title} {\bibinfo {title} {{Theory of
  Superconductivity}},\ }\href {https://doi.org/10.1103/PhysRev.108.1175}
  {\bibfield  {journal} {\bibinfo  {journal} {Phys. Rev.}\ }\textbf {\bibinfo
  {volume} {108}},\ \bibinfo {pages} {1175} (\bibinfo {year}
  {1957})}\BibitemShut {NoStop}%
\bibitem [{\citenamefont {Anderson}(1958)}]{anderson1958}%
  \BibitemOpen
  \bibfield  {author} {\bibinfo {author} {\bibfnamefont {P.~W.}\ \bibnamefont
  {Anderson}},\ }\bibfield  {title} {\bibinfo {title} {Absence of diffusion in
  certain random lattices},\ }\href {https://doi.org/10.1103/PhysRev.109.1492}
  {\bibfield  {journal} {\bibinfo  {journal} {Phys. Rev.}\ }\textbf {\bibinfo
  {volume} {109}},\ \bibinfo {pages} {1492} (\bibinfo {year}
  {1958})}\BibitemShut {NoStop}%
\bibitem [{\citenamefont {Hopfield}(1982)}]{hopfield1982}%
  \BibitemOpen
  \bibfield  {author} {\bibinfo {author} {\bibfnamefont {J.~J.}\ \bibnamefont
  {Hopfield}},\ }\bibfield  {title} {\bibinfo {title} {Neural networks and
  physical systems with emergent collective computational abilities.},\ }\href
  {https://doi.org/https://doi.org/10.1073/pnas.79.8.2554} {\bibfield
  {journal} {\bibinfo  {journal} {Proc. Natl. Acad. Sci. USA}\ }\textbf
  {\bibinfo {volume} {79}},\ \bibinfo {pages} {2554} (\bibinfo {year}
  {1982})}\BibitemShut {NoStop}%
\bibitem [{\citenamefont {Hinrichsen}(2000)}]{hinrichsen2000}%
  \BibitemOpen
  \bibfield  {author} {\bibinfo {author} {\bibfnamefont {H.}~\bibnamefont
  {Hinrichsen}},\ }\bibfield  {title} {\bibinfo {title} {{Non-equilibrium
  critical phenomena and phase transitions into absorbing states}},\ }\href
  {https://doi.org/10.1080/00018730050198152} {\bibfield  {journal} {\bibinfo
  {journal} {Adv. Phys.}\ }\textbf {\bibinfo {volume} {49}},\ \bibinfo {pages}
  {815–958} (\bibinfo {year} {2000})}\BibitemShut {NoStop}%
\bibitem [{\citenamefont {Landau}\ and\ \citenamefont
  {Lifshitz}(2013)}]{landau2013}%
  \BibitemOpen
  \bibfield  {author} {\bibinfo {author} {\bibfnamefont {L.~D.}\ \bibnamefont
  {Landau}}\ and\ \bibinfo {author} {\bibfnamefont {E.~M.}\ \bibnamefont
  {Lifshitz}},\ }\href@noop {} {\emph {\bibinfo {title} {Statistical Physics:
  Volume 5}}},\ Vol.~\bibinfo {volume} {5}\ (\bibinfo  {publisher} {Elsevier},\
  \bibinfo {year} {2013})\BibitemShut {NoStop}%
\bibitem [{\citenamefont {Sachdev}(2011)}]{sachdev2016}%
  \BibitemOpen
  \bibfield  {author} {\bibinfo {author} {\bibfnamefont {S.}~\bibnamefont
  {Sachdev}},\ }\href@noop {} {\emph {\bibinfo {title} {Quantum Phase
  Transitions}}}\ (\bibinfo  {publisher} {Cambridge University Press},\
  \bibinfo {year} {2011})\BibitemShut {NoStop}%
\bibitem [{\citenamefont {Huang}(2008)}]{huang2008}%
  \BibitemOpen
  \bibfield  {author} {\bibinfo {author} {\bibfnamefont {K.}~\bibnamefont
  {Huang}},\ }\href@noop {} {\emph {\bibinfo {title} {Statistical mechanics}}}\
  (\bibinfo  {publisher} {John Wiley \& Sons},\ \bibinfo {year}
  {2008})\BibitemShut {NoStop}%
\bibitem [{\citenamefont {Strocchi}(2005)}]{strocchi2005}%
  \BibitemOpen
  \bibfield  {author} {\bibinfo {author} {\bibfnamefont {F.}~\bibnamefont
  {Strocchi}},\ }\href@noop {} {\emph {\bibinfo {title} {Symmetry breaking}}},\
  Vol.\ \bibinfo {volume} {643}\ (\bibinfo  {publisher} {Springer},\ \bibinfo
  {year} {2005})\BibitemShut {NoStop}%
\bibitem [{\citenamefont {Itzykson}\ and\ \citenamefont
  {Drouffe}(1989)}]{itzykson1989}%
  \BibitemOpen
  \bibfield  {author} {\bibinfo {author} {\bibfnamefont {C.}~\bibnamefont
  {Itzykson}}\ and\ \bibinfo {author} {\bibfnamefont {J.-M.}\ \bibnamefont
  {Drouffe}},\ }\href@noop {} {\emph {\bibinfo {title} {Statistical Field
  Theory}}},\ Cambridge Monographs on Mathematical Physics\ (\bibinfo
  {publisher} {Cambridge University Press},\ \bibinfo {year}
  {1989})\BibitemShut {NoStop}%
\bibitem [{\citenamefont {Ruelle}(1969)}]{ruelle1969}%
  \BibitemOpen
  \bibfield  {author} {\bibinfo {author} {\bibfnamefont {D.}~\bibnamefont
  {Ruelle}},\ }\href@noop {} {\emph {\bibinfo {title} {Statistical mechanics:
  Rigorous results}}}\ (\bibinfo  {publisher} {World Scientific},\ \bibinfo
  {year} {1969})\BibitemShut {NoStop}%
\bibitem [{\citenamefont {Baxter}(2016)}]{baxter2016}%
  \BibitemOpen
  \bibfield  {author} {\bibinfo {author} {\bibfnamefont {R.~J.}\ \bibnamefont
  {Baxter}},\ }\href@noop {} {\emph {\bibinfo {title} {Exactly solved models in
  statistical mechanics}}}\ (\bibinfo  {publisher} {Elsevier},\ \bibinfo {year}
  {2016})\BibitemShut {NoStop}%
\bibitem [{\citenamefont {Ising}(1925)}]{ising1925}%
  \BibitemOpen
  \bibfield  {author} {\bibinfo {author} {\bibfnamefont {E.}~\bibnamefont
  {Ising}},\ }\bibfield  {title} {\bibinfo {title} {{Beitrag zur Theorie des
  Ferromagnetismus}},\ }\href {https://doi.org/10.1007/BF02980577} {\bibfield
  {journal} {\bibinfo  {journal} {Z. Phys.}\ }\textbf {\bibinfo {volume}
  {31}},\ \bibinfo {pages} {253} (\bibinfo {year} {1925})}\BibitemShut
  {NoStop}%
\bibitem [{\citenamefont {Onsager}(1944)}]{onsager1944}%
  \BibitemOpen
  \bibfield  {author} {\bibinfo {author} {\bibfnamefont {L.}~\bibnamefont
  {Onsager}},\ }\bibfield  {title} {\bibinfo {title} {Crystal statistics. i. a
  two-dimensional model with an order-disorder transition},\ }\href
  {https://doi.org/10.1103/PhysRev.65.117} {\bibfield  {journal} {\bibinfo
  {journal} {Phys. Rev.}\ }\textbf {\bibinfo {volume} {65}},\ \bibinfo {pages}
  {117} (\bibinfo {year} {1944})}\BibitemShut {NoStop}%
\bibitem [{\citenamefont {Goldenfeld}(2018)}]{goldenfeld2018}%
  \BibitemOpen
  \bibfield  {author} {\bibinfo {author} {\bibfnamefont {N.}~\bibnamefont
  {Goldenfeld}},\ }\href@noop {} {\emph {\bibinfo {title} {Lectures on phase
  transitions and the renormalization group}}}\ (\bibinfo  {publisher} {CRC
  Press},\ \bibinfo {year} {2018})\BibitemShut {NoStop}%
\bibitem [{\citenamefont {Davies}(1982)}]{davies1982}%
  \BibitemOpen
  \bibfield  {author} {\bibinfo {author} {\bibfnamefont {E.~B.}\ \bibnamefont
  {Davies}},\ }\bibfield  {title} {\bibinfo {title} {Metastability and the
  ising model},\ }\href {https://doi.org/10.1007/BF01013440} {\bibfield
  {journal} {\bibinfo  {journal} {J. Stat. Phys.}\ }\textbf {\bibinfo {volume}
  {27}},\ \bibinfo {pages} {657} (\bibinfo {year} {1982})}\BibitemShut
  {NoStop}%
\bibitem [{\citenamefont {McCraw}\ and\ \citenamefont
  {Schulman}(1978)}]{mccraw1978}%
  \BibitemOpen
  \bibfield  {author} {\bibinfo {author} {\bibfnamefont {R.}~\bibnamefont
  {McCraw}}\ and\ \bibinfo {author} {\bibfnamefont {L.}~\bibnamefont
  {Schulman}},\ }\bibfield  {title} {\bibinfo {title} {Metastability in the
  two-dimensional ising model},\ }\href {https://doi.org/10.1007/BF01018095}
  {\bibfield  {journal} {\bibinfo  {journal} {J. Stat. Phys.}\ }\textbf
  {\bibinfo {volume} {18}},\ \bibinfo {pages} {293} (\bibinfo {year}
  {1978})}\BibitemShut {NoStop}%
\bibitem [{\citenamefont {Mori}\ \emph {et~al.}(2018)\citenamefont {Mori},
  \citenamefont {Ikeda}, \citenamefont {Kaminishi},\ and\ \citenamefont
  {Ueda}}]{mori2018}%
  \BibitemOpen
  \bibfield  {author} {\bibinfo {author} {\bibfnamefont {T.}~\bibnamefont
  {Mori}}, \bibinfo {author} {\bibfnamefont {T.~N.}\ \bibnamefont {Ikeda}},
  \bibinfo {author} {\bibfnamefont {E.}~\bibnamefont {Kaminishi}},\ and\
  \bibinfo {author} {\bibfnamefont {M.}~\bibnamefont {Ueda}},\ }\bibfield
  {title} {\bibinfo {title} {Thermalization and prethermalization in isolated
  quantum systems: a theoretical overview},\ }\href
  {https://doi.org/10.1088/1361-6455/aabcdf} {\bibfield  {journal} {\bibinfo
  {journal} {J. Phys. B: At. Mol. Opt. Phys.}\ }\textbf {\bibinfo {volume}
  {51}},\ \bibinfo {pages} {112001} (\bibinfo {year} {2018})}\BibitemShut
  {NoStop}%
\bibitem [{\citenamefont {Yin}\ \emph {et~al.}(2025)\citenamefont {Yin},
  \citenamefont {Surace},\ and\ \citenamefont {Lucas}}]{yin2025}%
  \BibitemOpen
  \bibfield  {author} {\bibinfo {author} {\bibfnamefont {C.}~\bibnamefont
  {Yin}}, \bibinfo {author} {\bibfnamefont {F.~M.}\ \bibnamefont {Surace}},\
  and\ \bibinfo {author} {\bibfnamefont {A.}~\bibnamefont {Lucas}},\ }\bibfield
   {title} {\bibinfo {title} {Theory of metastable states in many-body quantum
  systems},\ }\href {https://doi.org/10.1103/PhysRevX.15.011064} {\bibfield
  {journal} {\bibinfo  {journal} {Phys. Rev. X}\ }\textbf {\bibinfo {volume}
  {15}},\ \bibinfo {pages} {011064} (\bibinfo {year} {2025})}\BibitemShut
  {NoStop}%
\bibitem [{\citenamefont {Macieszczak}\ \emph {et~al.}(2016)\citenamefont
  {Macieszczak}, \citenamefont {Guta}, \citenamefont {Lesanovsky},\ and\
  \citenamefont {Garrahan}}]{macieszczak2016}%
  \BibitemOpen
  \bibfield  {author} {\bibinfo {author} {\bibfnamefont {K.}~\bibnamefont
  {Macieszczak}}, \bibinfo {author} {\bibfnamefont {M.}~\bibnamefont {Guta}},
  \bibinfo {author} {\bibfnamefont {I.}~\bibnamefont {Lesanovsky}},\ and\
  \bibinfo {author} {\bibfnamefont {J.~P.}\ \bibnamefont {Garrahan}},\
  }\bibfield  {title} {\bibinfo {title} {{Towards a Theory of Metastability in
  Open Quantum Dynamics}},\ }\href
  {https://doi.org/10.1103/PhysRevLett.116.240404} {\bibfield  {journal}
  {\bibinfo  {journal} {Phys. Rev. Lett.}\ }\textbf {\bibinfo {volume} {116}},\
  \bibinfo {pages} {240404} (\bibinfo {year} {2016})}\BibitemShut {NoStop}%
\bibitem [{\citenamefont {Macieszczak}\ \emph {et~al.}(2021)\citenamefont
  {Macieszczak}, \citenamefont {Rose}, \citenamefont {Lesanovsky},\ and\
  \citenamefont {Garrahan}}]{macieszczak2021}%
  \BibitemOpen
  \bibfield  {author} {\bibinfo {author} {\bibfnamefont {K.}~\bibnamefont
  {Macieszczak}}, \bibinfo {author} {\bibfnamefont {D.~C.}\ \bibnamefont
  {Rose}}, \bibinfo {author} {\bibfnamefont {I.}~\bibnamefont {Lesanovsky}},\
  and\ \bibinfo {author} {\bibfnamefont {J.~P.}\ \bibnamefont {Garrahan}},\
  }\bibfield  {title} {\bibinfo {title} {Theory of classical metastability in
  open quantum systems},\ }\href
  {https://doi.org/10.1103/PhysRevResearch.3.033047} {\bibfield  {journal}
  {\bibinfo  {journal} {Phys. Rev. Res.}\ }\textbf {\bibinfo {volume} {3}},\
  \bibinfo {pages} {033047} (\bibinfo {year} {2021})}\BibitemShut {NoStop}%
\bibitem [{\citenamefont {M{\'e}zard}\ \emph {et~al.}(1987)\citenamefont
  {M{\'e}zard}, \citenamefont {Parisi},\ and\ \citenamefont
  {Virasoro}}]{mezard1987}%
  \BibitemOpen
  \bibfield  {author} {\bibinfo {author} {\bibfnamefont {M.}~\bibnamefont
  {M{\'e}zard}}, \bibinfo {author} {\bibfnamefont {G.}~\bibnamefont {Parisi}},\
  and\ \bibinfo {author} {\bibfnamefont {M.~A.}\ \bibnamefont {Virasoro}},\
  }\href@noop {} {\emph {\bibinfo {title} {Spin glass theory and beyond: An
  Introduction to the Replica Method and Its Applications}}},\ Vol.~\bibinfo
  {volume} {9}\ (\bibinfo  {publisher} {World Scientific Publishing Company},\
  \bibinfo {year} {1987})\BibitemShut {NoStop}%
\bibitem [{\citenamefont {Amit}\ \emph
  {et~al.}(1985{\natexlab{a}})\citenamefont {Amit}, \citenamefont {Gutfreund},\
  and\ \citenamefont {Sompolinsky}}]{amit1985a}%
  \BibitemOpen
  \bibfield  {author} {\bibinfo {author} {\bibfnamefont {D.~J.}\ \bibnamefont
  {Amit}}, \bibinfo {author} {\bibfnamefont {H.}~\bibnamefont {Gutfreund}},\
  and\ \bibinfo {author} {\bibfnamefont {H.}~\bibnamefont {Sompolinsky}},\
  }\bibfield  {title} {\bibinfo {title} {Spin-glass models of neural
  networks},\ }\href {https://doi.org/10.1103/PhysRevA.32.1007} {\bibfield
  {journal} {\bibinfo  {journal} {Phys. Rev. A}\ }\textbf {\bibinfo {volume}
  {32}},\ \bibinfo {pages} {1007} (\bibinfo {year}
  {1985}{\natexlab{a}})}\BibitemShut {NoStop}%
\bibitem [{\citenamefont {Amit}\ \emph
  {et~al.}(1985{\natexlab{b}})\citenamefont {Amit}, \citenamefont {Gutfreund},\
  and\ \citenamefont {Sompolinsky}}]{amit1985b}%
  \BibitemOpen
  \bibfield  {author} {\bibinfo {author} {\bibfnamefont {D.~J.}\ \bibnamefont
  {Amit}}, \bibinfo {author} {\bibfnamefont {H.}~\bibnamefont {Gutfreund}},\
  and\ \bibinfo {author} {\bibfnamefont {H.}~\bibnamefont {Sompolinsky}},\
  }\bibfield  {title} {\bibinfo {title} {{Storing Infinite Numbers of Patterns
  in a Spin-Glass Model of Neural Networks}},\ }\href
  {https://doi.org/10.1103/PhysRevLett.55.1530} {\bibfield  {journal} {\bibinfo
   {journal} {Phys. Rev. Lett.}\ }\textbf {\bibinfo {volume} {55}},\ \bibinfo
  {pages} {1530} (\bibinfo {year} {1985}{\natexlab{b}})}\BibitemShut {NoStop}%
\bibitem [{\citenamefont {Amit}\ \emph {et~al.}(1987)\citenamefont {Amit},
  \citenamefont {Gutfreund},\ and\ \citenamefont {Sompolinsky}}]{amit1987}%
  \BibitemOpen
  \bibfield  {author} {\bibinfo {author} {\bibfnamefont {D.~J.}\ \bibnamefont
  {Amit}}, \bibinfo {author} {\bibfnamefont {H.}~\bibnamefont {Gutfreund}},\
  and\ \bibinfo {author} {\bibfnamefont {H.}~\bibnamefont {Sompolinsky}},\
  }\bibfield  {title} {\bibinfo {title} {Statistical mechanics of neural
  networks near saturation},\ }\href
  {https://doi.org/https://doi.org/10.1016/0003-4916(87)90092-3} {\bibfield
  {journal} {\bibinfo  {journal} {Ann. Phys.}\ }\textbf {\bibinfo {volume}
  {173}},\ \bibinfo {pages} {30} (\bibinfo {year} {1987})}\BibitemShut
  {NoStop}%
\bibitem [{\citenamefont {Biamonte}\ \emph {et~al.}(2017)\citenamefont
  {Biamonte}, \citenamefont {Wittek}, \citenamefont {Pancotti}, \citenamefont
  {Rebentrost}, \citenamefont {Wiebe},\ and\ \citenamefont
  {Lloyd}}]{biamonte2017}%
  \BibitemOpen
  \bibfield  {author} {\bibinfo {author} {\bibfnamefont {J.}~\bibnamefont
  {Biamonte}}, \bibinfo {author} {\bibfnamefont {P.}~\bibnamefont {Wittek}},
  \bibinfo {author} {\bibfnamefont {N.}~\bibnamefont {Pancotti}}, \bibinfo
  {author} {\bibfnamefont {P.}~\bibnamefont {Rebentrost}}, \bibinfo {author}
  {\bibfnamefont {N.}~\bibnamefont {Wiebe}},\ and\ \bibinfo {author}
  {\bibfnamefont {S.}~\bibnamefont {Lloyd}},\ }\bibfield  {title} {\bibinfo
  {title} {Quantum machine learning},\ }\href
  {https://doi.org/https://doi.org/10.1038/nature23474} {\bibfield  {journal}
  {\bibinfo  {journal} {Nature}\ }\textbf {\bibinfo {volume} {549}},\ \bibinfo
  {pages} {195} (\bibinfo {year} {2017})}\BibitemShut {NoStop}%
\bibitem [{\citenamefont {Schuld}\ and\ \citenamefont
  {Petruccione}(2021)}]{schuld2021}%
  \BibitemOpen
  \bibfield  {author} {\bibinfo {author} {\bibfnamefont {M.}~\bibnamefont
  {Schuld}}\ and\ \bibinfo {author} {\bibfnamefont {F.}~\bibnamefont
  {Petruccione}},\ }\href@noop {} {\emph {\bibinfo {title} {{Machine Learning
  with Quantum Computers}}}}\ (\bibinfo  {publisher} {Springer},\ \bibinfo
  {year} {2021})\BibitemShut {NoStop}%
\bibitem [{\citenamefont {Schuld}\ and\ \citenamefont
  {Killoran}(2022)}]{schuld2022}%
  \BibitemOpen
  \bibfield  {author} {\bibinfo {author} {\bibfnamefont {M.}~\bibnamefont
  {Schuld}}\ and\ \bibinfo {author} {\bibfnamefont {N.}~\bibnamefont
  {Killoran}},\ }\bibfield  {title} {\bibinfo {title} {{Is Quantum Advantage
  the Right Goal for Quantum Machine Learning?}},\ }\href
  {https://doi.org/10.1103/PRXQuantum.3.030101} {\bibfield  {journal} {\bibinfo
   {journal} {PRX Quantum}\ }\textbf {\bibinfo {volume} {3}},\ \bibinfo {pages}
  {030101} (\bibinfo {year} {2022})}\BibitemShut {NoStop}%
\bibitem [{\citenamefont {Cerezo}\ \emph {et~al.}(2022)\citenamefont {Cerezo},
  \citenamefont {Verdon}, \citenamefont {Huang}, \citenamefont {Cincio},\ and\
  \citenamefont {Coles}}]{cerezo2022}%
  \BibitemOpen
  \bibfield  {author} {\bibinfo {author} {\bibfnamefont {M.}~\bibnamefont
  {Cerezo}}, \bibinfo {author} {\bibfnamefont {G.}~\bibnamefont {Verdon}},
  \bibinfo {author} {\bibfnamefont {H.-Y.}\ \bibnamefont {Huang}}, \bibinfo
  {author} {\bibfnamefont {L.}~\bibnamefont {Cincio}},\ and\ \bibinfo {author}
  {\bibfnamefont {P.~J.}\ \bibnamefont {Coles}},\ }\bibfield  {title} {\bibinfo
  {title} {Challenges and opportunities in quantum machine learning},\ }\href
  {https://doi.org/10.1038/s43588-022-00311-3} {\bibfield  {journal} {\bibinfo
  {journal} {Nat. Comput. Sci.}\ }\textbf {\bibinfo {volume} {2}},\ \bibinfo
  {pages} {567} (\bibinfo {year} {2022})}\BibitemShut {NoStop}%
\bibitem [{\citenamefont {Farrelly}(2020)}]{farrelly2020}%
  \BibitemOpen
  \bibfield  {author} {\bibinfo {author} {\bibfnamefont {T.}~\bibnamefont
  {Farrelly}},\ }\bibfield  {title} {\bibinfo {title} {A review of {Q}uantum
  {C}ellular {A}utomata},\ }\href {https://doi.org/10.22331/q-2020-11-30-368}
  {\bibfield  {journal} {\bibinfo  {journal} {{Quantum}}\ }\textbf {\bibinfo
  {volume} {4}},\ \bibinfo {pages} {368} (\bibinfo {year} {2020})}\BibitemShut
  {NoStop}%
\bibitem [{\citenamefont {Arrighi}(2019)}]{arrighi2019}%
  \BibitemOpen
  \bibfield  {author} {\bibinfo {author} {\bibfnamefont {P.}~\bibnamefont
  {Arrighi}},\ }\bibfield  {title} {\bibinfo {title} {An overview of quantum
  cellular automata},\ }\href
  {https://doi.org/https://doi.org/10.1007/s11047-019-09762-6} {\bibfield
  {journal} {\bibinfo  {journal} {Nat. Comput.}\ }\textbf {\bibinfo {volume}
  {18}},\ \bibinfo {pages} {885} (\bibinfo {year} {2019})}\BibitemShut
  {NoStop}%
\bibitem [{\citenamefont {Lesanovsky}\ \emph {et~al.}(2019)\citenamefont
  {Lesanovsky}, \citenamefont {Macieszczak},\ and\ \citenamefont
  {Garrahan}}]{lesanovsky2019}%
  \BibitemOpen
  \bibfield  {author} {\bibinfo {author} {\bibfnamefont {I.}~\bibnamefont
  {Lesanovsky}}, \bibinfo {author} {\bibfnamefont {K.}~\bibnamefont
  {Macieszczak}},\ and\ \bibinfo {author} {\bibfnamefont {J.~P.}\ \bibnamefont
  {Garrahan}},\ }\bibfield  {title} {\bibinfo {title} {Non-equilibrium
  absorbing state phase transitions in discrete-time quantum cellular automaton
  dynamics on spin lattices},\ }\href
  {https://doi.org/10.1088/2058-9565/aaf831} {\bibfield  {journal} {\bibinfo
  {journal} {Quantum Sci. Technol.}\ }\textbf {\bibinfo {volume} {4}},\
  \bibinfo {pages} {02LT02} (\bibinfo {year} {2019})}\BibitemShut {NoStop}%
\bibitem [{\citenamefont {Gillman}\ \emph {et~al.}(2020)\citenamefont
  {Gillman}, \citenamefont {Carollo},\ and\ \citenamefont
  {Lesanovsky}}]{gillman2020}%
  \BibitemOpen
  \bibfield  {author} {\bibinfo {author} {\bibfnamefont {E.}~\bibnamefont
  {Gillman}}, \bibinfo {author} {\bibfnamefont {F.}~\bibnamefont {Carollo}},\
  and\ \bibinfo {author} {\bibfnamefont {I.}~\bibnamefont {Lesanovsky}},\
  }\bibfield  {title} {\bibinfo {title} {{Nonequilibrium Phase Transitions in
  ($1+1$)-Dimensional Quantum Cellular Automata with Controllable Quantum
  Correlations}},\ }\href {https://doi.org/10.1103/PhysRevLett.125.100403}
  {\bibfield  {journal} {\bibinfo  {journal} {Phys. Rev. Lett.}\ }\textbf
  {\bibinfo {volume} {125}},\ \bibinfo {pages} {100403} (\bibinfo {year}
  {2020})}\BibitemShut {NoStop}%
\bibitem [{\citenamefont {Gillman}\ \emph
  {et~al.}(2021{\natexlab{a}})\citenamefont {Gillman}, \citenamefont
  {Carollo},\ and\ \citenamefont {Lesanovsky}}]{gillman2021a}%
  \BibitemOpen
  \bibfield  {author} {\bibinfo {author} {\bibfnamefont {E.}~\bibnamefont
  {Gillman}}, \bibinfo {author} {\bibfnamefont {F.}~\bibnamefont {Carollo}},\
  and\ \bibinfo {author} {\bibfnamefont {I.}~\bibnamefont {Lesanovsky}},\
  }\bibfield  {title} {\bibinfo {title} {Numerical simulation of quantum
  nonequilibrium phase transitions without finite-size effects},\ }\href
  {https://doi.org/10.1103/PhysRevA.103.L040201} {\bibfield  {journal}
  {\bibinfo  {journal} {Phys. Rev. A}\ }\textbf {\bibinfo {volume} {103}},\
  \bibinfo {pages} {L040201} (\bibinfo {year}
  {2021}{\natexlab{a}})}\BibitemShut {NoStop}%
\bibitem [{\citenamefont {Gillman}\ \emph
  {et~al.}(2021{\natexlab{b}})\citenamefont {Gillman}, \citenamefont
  {Carollo},\ and\ \citenamefont {Lesanovsky}}]{gillman2021b}%
  \BibitemOpen
  \bibfield  {author} {\bibinfo {author} {\bibfnamefont {E.}~\bibnamefont
  {Gillman}}, \bibinfo {author} {\bibfnamefont {F.}~\bibnamefont {Carollo}},\
  and\ \bibinfo {author} {\bibfnamefont {I.}~\bibnamefont {Lesanovsky}},\
  }\bibfield  {title} {\bibinfo {title} {{Quantum and Classical Temporal
  Correlations in $(1+1)\mathrm{D}$ Quantum Cellular Automata}},\ }\href
  {https://doi.org/10.1103/PhysRevLett.127.230502} {\bibfield  {journal}
  {\bibinfo  {journal} {Phys. Rev. Lett.}\ }\textbf {\bibinfo {volume} {127}},\
  \bibinfo {pages} {230502} (\bibinfo {year} {2021}{\natexlab{b}})}\BibitemShut
  {NoStop}%
\bibitem [{\citenamefont {Nigmatullin}\ \emph {et~al.}(2021)\citenamefont
  {Nigmatullin}, \citenamefont {Wagner},\ and\ \citenamefont
  {Brennen}}]{nigmatullin2021}%
  \BibitemOpen
  \bibfield  {author} {\bibinfo {author} {\bibfnamefont {R.}~\bibnamefont
  {Nigmatullin}}, \bibinfo {author} {\bibfnamefont {E.}~\bibnamefont
  {Wagner}},\ and\ \bibinfo {author} {\bibfnamefont {G.~K.}\ \bibnamefont
  {Brennen}},\ }\bibfield  {title} {\bibinfo {title} {Directed percolation in
  nonunitary quantum cellular automata},\ }\href
  {https://doi.org/10.1103/PhysRevResearch.3.043167} {\bibfield  {journal}
  {\bibinfo  {journal} {Phys. Rev. Res.}\ }\textbf {\bibinfo {volume} {3}},\
  \bibinfo {pages} {043167} (\bibinfo {year} {2021})}\BibitemShut {NoStop}%
\bibitem [{\citenamefont {Gillman}\ \emph {et~al.}(2022)\citenamefont
  {Gillman}, \citenamefont {Carollo},\ and\ \citenamefont
  {Lesanovsky}}]{gillman2022a}%
  \BibitemOpen
  \bibfield  {author} {\bibinfo {author} {\bibfnamefont {E.}~\bibnamefont
  {Gillman}}, \bibinfo {author} {\bibfnamefont {F.}~\bibnamefont {Carollo}},\
  and\ \bibinfo {author} {\bibfnamefont {I.}~\bibnamefont {Lesanovsky}},\
  }\bibfield  {title} {\bibinfo {title} {{Asynchronism and nonequilibrium phase
  transitions in $(1+1)$-dimensional quantum cellular automata}},\ }\href
  {https://doi.org/10.1103/PhysRevE.106.L032103} {\bibfield  {journal}
  {\bibinfo  {journal} {Phys. Rev. E}\ }\textbf {\bibinfo {volume} {106}},\
  \bibinfo {pages} {L032103} (\bibinfo {year} {2022})}\BibitemShut {NoStop}%
\bibitem [{\citenamefont {Guedes}\ \emph {et~al.}(2024)\citenamefont {Guedes},
  \citenamefont {Winter},\ and\ \citenamefont {M\"uller}}]{guedes2024}%
  \BibitemOpen
  \bibfield  {author} {\bibinfo {author} {\bibfnamefont {T.~L.~M.}\
  \bibnamefont {Guedes}}, \bibinfo {author} {\bibfnamefont {D.}~\bibnamefont
  {Winter}},\ and\ \bibinfo {author} {\bibfnamefont {M.}~\bibnamefont
  {M\"uller}},\ }\bibfield  {title} {\bibinfo {title} {Quantum cellular
  automata for quantum error correction and density classification},\ }\href
  {https://doi.org/10.1103/PhysRevLett.133.150601} {\bibfield  {journal}
  {\bibinfo  {journal} {Phys. Rev. Lett.}\ }\textbf {\bibinfo {volume} {133}},\
  \bibinfo {pages} {150601} (\bibinfo {year} {2024})}\BibitemShut {NoStop}%
\bibitem [{\citenamefont {Wagner}\ \emph {et~al.}(2024)\citenamefont {Wagner},
  \citenamefont {Nigmatullin}, \citenamefont {Gilchrist},\ and\ \citenamefont
  {Brennen}}]{wagner2024}%
  \BibitemOpen
  \bibfield  {author} {\bibinfo {author} {\bibfnamefont {E.}~\bibnamefont
  {Wagner}}, \bibinfo {author} {\bibfnamefont {R.}~\bibnamefont {Nigmatullin}},
  \bibinfo {author} {\bibfnamefont {A.}~\bibnamefont {Gilchrist}},\ and\
  \bibinfo {author} {\bibfnamefont {G.~K.}\ \bibnamefont {Brennen}},\
  }\bibfield  {title} {\bibinfo {title} {{Information flow in non-unitary
  quantum cellular automata}},\ }\href
  {https://doi.org/10.21468/SciPostPhys.16.1.014} {\bibfield  {journal}
  {\bibinfo  {journal} {SciPost Phys.}\ }\textbf {\bibinfo {volume} {16}},\
  \bibinfo {pages} {014} (\bibinfo {year} {2024})}\BibitemShut {NoStop}%
\bibitem [{\citenamefont {Wagner}\ \emph {et~al.}(2025)\citenamefont {Wagner},
  \citenamefont {Dell’Anna}, \citenamefont {Nigmatullin},\ and\ \citenamefont
  {K.~Brennen}}]{wagner2025}%
  \BibitemOpen
  \bibfield  {author} {\bibinfo {author} {\bibfnamefont {E.}~\bibnamefont
  {Wagner}}, \bibinfo {author} {\bibfnamefont {F.}~\bibnamefont {Dell’Anna}},
  \bibinfo {author} {\bibfnamefont {R.}~\bibnamefont {Nigmatullin}},\ and\
  \bibinfo {author} {\bibfnamefont {G.}~\bibnamefont {K.~Brennen}},\ }\bibfield
   {title} {\bibinfo {title} {Density classification with non-unitary quantum
  cellular automata},\ }\href {https://doi.org/10.3390/e27010026} {\bibfield
  {journal} {\bibinfo  {journal} {Entropy}\ }\textbf {\bibinfo {volume} {27}}
  (\bibinfo {year} {2025})}\BibitemShut {NoStop}%
\bibitem [{\citenamefont {Beer}\ \emph {et~al.}(2020)\citenamefont {Beer},
  \citenamefont {Bondarenko}, \citenamefont {Farrelly}, \citenamefont
  {Osborne}, \citenamefont {Salzmann}, \citenamefont {Scheiermann},\ and\
  \citenamefont {Wolf}}]{beer2020}%
  \BibitemOpen
  \bibfield  {author} {\bibinfo {author} {\bibfnamefont {K.}~\bibnamefont
  {Beer}}, \bibinfo {author} {\bibfnamefont {D.}~\bibnamefont {Bondarenko}},
  \bibinfo {author} {\bibfnamefont {T.}~\bibnamefont {Farrelly}}, \bibinfo
  {author} {\bibfnamefont {T.~J.}\ \bibnamefont {Osborne}}, \bibinfo {author}
  {\bibfnamefont {R.}~\bibnamefont {Salzmann}}, \bibinfo {author}
  {\bibfnamefont {D.}~\bibnamefont {Scheiermann}},\ and\ \bibinfo {author}
  {\bibfnamefont {R.}~\bibnamefont {Wolf}},\ }\bibfield  {title} {\bibinfo
  {title} {Training deep quantum neural networks},\ }\href
  {https://doi.org/https://doi.org/10.1038/s41467-020-14454-2} {\bibfield
  {journal} {\bibinfo  {journal} {Nat. Commun.}\ }\textbf {\bibinfo {volume}
  {11}},\ \bibinfo {pages} {1} (\bibinfo {year} {2020})}\BibitemShut {NoStop}%
\bibitem [{\citenamefont {Bondarenko}\ and\ \citenamefont
  {Feldmann}(2020)}]{bondarenko2020}%
  \BibitemOpen
  \bibfield  {author} {\bibinfo {author} {\bibfnamefont {D.}~\bibnamefont
  {Bondarenko}}\ and\ \bibinfo {author} {\bibfnamefont {P.}~\bibnamefont
  {Feldmann}},\ }\bibfield  {title} {\bibinfo {title} {Quantum autoencoders to
  denoise quantum data},\ }\href
  {https://link.aps.org/doi/10.1103/PhysRevLett.124.130502} {\bibfield
  {journal} {\bibinfo  {journal} {Phys. Rev. Lett.}\ }\textbf {\bibinfo
  {volume} {124}},\ \bibinfo {pages} {130502} (\bibinfo {year}
  {2020})}\BibitemShut {NoStop}%
\bibitem [{\citenamefont {Lewenstein}\ \emph {et~al.}(2021)\citenamefont
  {Lewenstein}, \citenamefont {Gratsea}, \citenamefont {Riera-Campeny},
  \citenamefont {Aloy}, \citenamefont {Kasper},\ and\ \citenamefont
  {Sanpera}}]{lewenstein2021}%
  \BibitemOpen
  \bibfield  {author} {\bibinfo {author} {\bibfnamefont {M.}~\bibnamefont
  {Lewenstein}}, \bibinfo {author} {\bibfnamefont {A.}~\bibnamefont {Gratsea}},
  \bibinfo {author} {\bibfnamefont {A.}~\bibnamefont {Riera-Campeny}}, \bibinfo
  {author} {\bibfnamefont {A.}~\bibnamefont {Aloy}}, \bibinfo {author}
  {\bibfnamefont {V.}~\bibnamefont {Kasper}},\ and\ \bibinfo {author}
  {\bibfnamefont {A.}~\bibnamefont {Sanpera}},\ }\bibfield  {title} {\bibinfo
  {title} {Storage capacity and learning capability of quantum neural
  networks},\ }\href {https://doi.org/10.1088/2058-9565/ac070f} {\bibfield
  {journal} {\bibinfo  {journal} {Quantum Sci. Technol.}\ }\textbf {\bibinfo
  {volume} {6}},\ \bibinfo {pages} {045002} (\bibinfo {year}
  {2021})}\BibitemShut {NoStop}%
\bibitem [{\citenamefont {Beer}(2022)}]{beer2022}%
  \BibitemOpen
  \bibfield  {author} {\bibinfo {author} {\bibfnamefont {K.}~\bibnamefont
  {Beer}},\ }\href {https://doi.org/10.15488/11896} {\bibinfo {title} {Quantum
  neural networks}} (\bibinfo {year} {2022})\BibitemShut {NoStop}%
\bibitem [{\citenamefont {Gillman}\ \emph {et~al.}(2023)\citenamefont
  {Gillman}, \citenamefont {Carollo},\ and\ \citenamefont
  {Lesanovsky}}]{gillman2022b}%
  \BibitemOpen
  \bibfield  {author} {\bibinfo {author} {\bibfnamefont {E.}~\bibnamefont
  {Gillman}}, \bibinfo {author} {\bibfnamefont {F.}~\bibnamefont {Carollo}},\
  and\ \bibinfo {author} {\bibfnamefont {I.}~\bibnamefont {Lesanovsky}},\
  }\bibfield  {title} {\bibinfo {title} {{Using $(1+1)D$ quantum cellular
  automata for exploring collective effects in large-scale quantum neural
  networks}},\ }\href {https://doi.org/10.1103/PhysRevE.107.L022102} {\bibfield
   {journal} {\bibinfo  {journal} {Phys. Rev. E}\ }\textbf {\bibinfo {volume}
  {107}},\ \bibinfo {pages} {L022102} (\bibinfo {year} {2023})}\BibitemShut
  {NoStop}%
\bibitem [{\citenamefont {Boneberg}\ \emph {et~al.}(2023)\citenamefont
  {Boneberg}, \citenamefont {Carollo},\ and\ \citenamefont
  {Lesanovsky}}]{boneberg2023}%
  \BibitemOpen
  \bibfield  {author} {\bibinfo {author} {\bibfnamefont {M.}~\bibnamefont
  {Boneberg}}, \bibinfo {author} {\bibfnamefont {F.}~\bibnamefont {Carollo}},\
  and\ \bibinfo {author} {\bibfnamefont {I.}~\bibnamefont {Lesanovsky}},\
  }\bibfield  {title} {\bibinfo {title} {{Dissipative quantum many-body
  dynamics in (1+ 1) D quantum cellular automata and quantum neural
  networks}},\ }\href {https://doi.org/10.1088/1367-2630/aceff4} {\bibfield
  {journal} {\bibinfo  {journal} {New J. Phys.}\ }\textbf {\bibinfo {volume}
  {25}},\ \bibinfo {pages} {093020} (\bibinfo {year} {2023})}\BibitemShut
  {NoStop}%
\bibitem [{\citenamefont {Locher}\ \emph {et~al.}(2023)\citenamefont {Locher},
  \citenamefont {Cardarelli},\ and\ \citenamefont {M{\"u}ller}}]{locher2023}%
  \BibitemOpen
  \bibfield  {author} {\bibinfo {author} {\bibfnamefont {D.~F.}\ \bibnamefont
  {Locher}}, \bibinfo {author} {\bibfnamefont {L.}~\bibnamefont {Cardarelli}},\
  and\ \bibinfo {author} {\bibfnamefont {M.}~\bibnamefont {M{\"u}ller}},\
  }\bibfield  {title} {\bibinfo {title} {Quantum error correction with quantum
  autoencoders},\ }\href {https://doi.org/10.22331/q-2023-03-09-942} {\bibfield
   {journal} {\bibinfo  {journal} {Quantum}\ }\textbf {\bibinfo {volume} {7}},\
  \bibinfo {pages} {942} (\bibinfo {year} {2023})}\BibitemShut {NoStop}%
\bibitem [{\citenamefont {Sutter}\ \emph {et~al.}(2025)\citenamefont {Sutter},
  \citenamefont {Popp},\ and\ \citenamefont {Hiesmayr}}]{sutter2025}%
  \BibitemOpen
  \bibfield  {author} {\bibinfo {author} {\bibfnamefont {T.~C.}\ \bibnamefont
  {Sutter}}, \bibinfo {author} {\bibfnamefont {C.}~\bibnamefont {Popp}},\ and\
  \bibinfo {author} {\bibfnamefont {B.~C.}\ \bibnamefont {Hiesmayr}},\
  }\bibfield  {title} {\bibinfo {title} {The impact of architecture and cost
  function on dissipative quantum neural networks},\ }\href
  {https://doi.org/10.48550/arXiv.2502.09526} {\bibfield  {journal} {\bibinfo
  {journal} {arXiv:2502.09526}\ } (\bibinfo {year} {2025})}\BibitemShut
  {NoStop}%
\bibitem [{\citenamefont {Boneberg}\ \emph {et~al.}(2025)\citenamefont
  {Boneberg}, \citenamefont {Carollo},\ and\ \citenamefont
  {Lesanovsky}}]{boneberg2025}%
  \BibitemOpen
  \bibfield  {author} {\bibinfo {author} {\bibfnamefont {M.}~\bibnamefont
  {Boneberg}}, \bibinfo {author} {\bibfnamefont {F.}~\bibnamefont {Carollo}},\
  and\ \bibinfo {author} {\bibfnamefont {I.}~\bibnamefont {Lesanovsky}},\
  }\bibfield  {title} {\bibinfo {title} {Nonlinear classification capability of
  quantum neural networks due to emergent quantum metastability},\ }\href
  {https://doi.org/10.1103/PhysRevA.111.062405} {\bibfield  {journal} {\bibinfo
   {journal} {Phys. Rev. A}\ }\textbf {\bibinfo {volume} {111}},\ \bibinfo
  {pages} {062405} (\bibinfo {year} {2025})}\BibitemShut {NoStop}%
\bibitem [{\citenamefont {Breuer}\ and\ \citenamefont
  {Petruccione}(2002)}]{breuer2002}%
  \BibitemOpen
  \bibfield  {author} {\bibinfo {author} {\bibfnamefont {H.-P.}\ \bibnamefont
  {Breuer}}\ and\ \bibinfo {author} {\bibfnamefont {F.}~\bibnamefont
  {Petruccione}},\ }\href@noop {} {\emph {\bibinfo {title} {The Theory of Open
  Quantum Systems}}}\ (\bibinfo  {publisher} {Oxford University Press},\
  \bibinfo {year} {2002})\BibitemShut {NoStop}%
\bibitem [{\citenamefont {Rivas}\ and\ \citenamefont
  {Huelga}(2012)}]{rivas2012}%
  \BibitemOpen
  \bibfield  {author} {\bibinfo {author} {\bibfnamefont {A.}~\bibnamefont
  {Rivas}}\ and\ \bibinfo {author} {\bibfnamefont {S.~F.}\ \bibnamefont
  {Huelga}},\ }\href@noop {} {\emph {\bibinfo {title} {Open quantum
  systems}}},\ Vol.~\bibinfo {volume} {10}\ (\bibinfo  {publisher} {Springer},\
  \bibinfo {year} {2012})\BibitemShut {NoStop}%
\bibitem [{\citenamefont {Bartolo}\ \emph {et~al.}(2016)\citenamefont
  {Bartolo}, \citenamefont {Minganti}, \citenamefont {Casteels},\ and\
  \citenamefont {Ciuti}}]{bartolo2016}%
  \BibitemOpen
  \bibfield  {author} {\bibinfo {author} {\bibfnamefont {N.}~\bibnamefont
  {Bartolo}}, \bibinfo {author} {\bibfnamefont {F.}~\bibnamefont {Minganti}},
  \bibinfo {author} {\bibfnamefont {W.}~\bibnamefont {Casteels}},\ and\
  \bibinfo {author} {\bibfnamefont {C.}~\bibnamefont {Ciuti}},\ }\bibfield
  {title} {\bibinfo {title} {Exact steady state of a kerr resonator with one-
  and two-photon driving and dissipation: Controllable wigner-function
  multimodality and dissipative phase transitions},\ }\href
  {https://doi.org/10.1103/PhysRevA.94.033841} {\bibfield  {journal} {\bibinfo
  {journal} {Phys. Rev. A}\ }\textbf {\bibinfo {volume} {94}},\ \bibinfo
  {pages} {033841} (\bibinfo {year} {2016})}\BibitemShut {NoStop}%
\bibitem [{\citenamefont {Minganti}\ \emph {et~al.}(2023)\citenamefont
  {Minganti}, \citenamefont {Savona},\ and\ \citenamefont
  {Biella}}]{minganti2023}%
  \BibitemOpen
  \bibfield  {author} {\bibinfo {author} {\bibfnamefont {F.}~\bibnamefont
  {Minganti}}, \bibinfo {author} {\bibfnamefont {V.}~\bibnamefont {Savona}},\
  and\ \bibinfo {author} {\bibfnamefont {A.}~\bibnamefont {Biella}},\
  }\bibfield  {title} {\bibinfo {title} {Dissipative phase transitions in
  {$n$}-photon driven quantum nonlinear resonators},\ }\href
  {https://doi.org/10.22331/q-2023-11-07-1170} {\bibfield  {journal} {\bibinfo
  {journal} {{Quantum}}\ }\textbf {\bibinfo {volume} {7}},\ \bibinfo {pages}
  {1170} (\bibinfo {year} {2023})}\BibitemShut {NoStop}%
\bibitem [{\citenamefont {Lieu}\ \emph {et~al.}(2020)\citenamefont {Lieu},
  \citenamefont {Belyansky}, \citenamefont {Young}, \citenamefont {Lundgren},
  \citenamefont {Albert},\ and\ \citenamefont {Gorshkov}}]{lieu2020}%
  \BibitemOpen
  \bibfield  {author} {\bibinfo {author} {\bibfnamefont {S.}~\bibnamefont
  {Lieu}}, \bibinfo {author} {\bibfnamefont {R.}~\bibnamefont {Belyansky}},
  \bibinfo {author} {\bibfnamefont {J.~T.}\ \bibnamefont {Young}}, \bibinfo
  {author} {\bibfnamefont {R.}~\bibnamefont {Lundgren}}, \bibinfo {author}
  {\bibfnamefont {V.~V.}\ \bibnamefont {Albert}},\ and\ \bibinfo {author}
  {\bibfnamefont {A.~V.}\ \bibnamefont {Gorshkov}},\ }\bibfield  {title}
  {\bibinfo {title} {Symmetry breaking and error correction in open quantum
  systems},\ }\href {https://doi.org/10.1103/PhysRevLett.125.240405} {\bibfield
   {journal} {\bibinfo  {journal} {Phys. Rev. Lett.}\ }\textbf {\bibinfo
  {volume} {125}},\ \bibinfo {pages} {240405} (\bibinfo {year}
  {2020})}\BibitemShut {NoStop}%
\bibitem [{\citenamefont {Bu{\v{c}}a}\ and\ \citenamefont
  {Prosen}(2012)}]{buca2012}%
  \BibitemOpen
  \bibfield  {author} {\bibinfo {author} {\bibfnamefont {B.}~\bibnamefont
  {Bu{\v{c}}a}}\ and\ \bibinfo {author} {\bibfnamefont {T.}~\bibnamefont
  {Prosen}},\ }\bibfield  {title} {\bibinfo {title} {A note on symmetry
  reductions of the lindblad equation: transport in constrained open spin
  chains},\ }\href {https://doi.org/10.1088/1367-2630/14/7/073007} {\bibfield
  {journal} {\bibinfo  {journal} {New J. Phys.}\ }\textbf {\bibinfo {volume}
  {14}},\ \bibinfo {pages} {073007} (\bibinfo {year} {2012})}\BibitemShut
  {NoStop}%
\bibitem [{\citenamefont {Albert}\ and\ \citenamefont
  {Jiang}(2014)}]{albert2014}%
  \BibitemOpen
  \bibfield  {author} {\bibinfo {author} {\bibfnamefont {V.~V.}\ \bibnamefont
  {Albert}}\ and\ \bibinfo {author} {\bibfnamefont {L.}~\bibnamefont {Jiang}},\
  }\bibfield  {title} {\bibinfo {title} {Symmetries and conserved quantities in
  lindblad master equations},\ }\href
  {https://doi.org/10.1103/PhysRevA.89.022118} {\bibfield  {journal} {\bibinfo
  {journal} {Phys. Rev. A}\ }\textbf {\bibinfo {volume} {89}},\ \bibinfo
  {pages} {022118} (\bibinfo {year} {2014})}\BibitemShut {NoStop}%
\bibitem [{\citenamefont {Sieberer}\ \emph {et~al.}(2025)\citenamefont
  {Sieberer}, \citenamefont {Buchhold}, \citenamefont {Marino},\ and\
  \citenamefont {Diehl}}]{sieberer2025}%
  \BibitemOpen
  \bibfield  {author} {\bibinfo {author} {\bibfnamefont {L.~M.}\ \bibnamefont
  {Sieberer}}, \bibinfo {author} {\bibfnamefont {M.}~\bibnamefont {Buchhold}},
  \bibinfo {author} {\bibfnamefont {J.}~\bibnamefont {Marino}},\ and\ \bibinfo
  {author} {\bibfnamefont {S.}~\bibnamefont {Diehl}},\ }\bibfield  {title}
  {\bibinfo {title} {Universality in driven open quantum matter},\ }\href
  {https://doi.org/10.1103/RevModPhys.97.025004} {\bibfield  {journal}
  {\bibinfo  {journal} {Rev. Mod. Phys.}\ }\textbf {\bibinfo {volume} {97}},\
  \bibinfo {pages} {025004} (\bibinfo {year} {2025})}\BibitemShut {NoStop}%
\bibitem [{\citenamefont {Overbeck}\ \emph {et~al.}(2017)\citenamefont
  {Overbeck}, \citenamefont {Maghrebi}, \citenamefont {Gorshkov},\ and\
  \citenamefont {Weimer}}]{overbeck2017}%
  \BibitemOpen
  \bibfield  {author} {\bibinfo {author} {\bibfnamefont {V.~R.}\ \bibnamefont
  {Overbeck}}, \bibinfo {author} {\bibfnamefont {M.~F.}\ \bibnamefont
  {Maghrebi}}, \bibinfo {author} {\bibfnamefont {A.~V.}\ \bibnamefont
  {Gorshkov}},\ and\ \bibinfo {author} {\bibfnamefont {H.}~\bibnamefont
  {Weimer}},\ }\bibfield  {title} {\bibinfo {title} {{Multicritical behavior in
  dissipative Ising models}},\ }\href
  {https://doi.org/10.1103/PhysRevA.95.042133} {\bibfield  {journal} {\bibinfo
  {journal} {Phys. Rev. A}\ }\textbf {\bibinfo {volume} {95}},\ \bibinfo
  {pages} {042133} (\bibinfo {year} {2017})}\BibitemShut {NoStop}%
\bibitem [{\citenamefont {McClean}\ \emph {et~al.}(2018)\citenamefont
  {McClean}, \citenamefont {Boixo}, \citenamefont {Smelyanskiy}, \citenamefont
  {Babbush},\ and\ \citenamefont {Neven}}]{mcclean2018}%
  \BibitemOpen
  \bibfield  {author} {\bibinfo {author} {\bibfnamefont {J.~R.}\ \bibnamefont
  {McClean}}, \bibinfo {author} {\bibfnamefont {S.}~\bibnamefont {Boixo}},
  \bibinfo {author} {\bibfnamefont {V.~N.}\ \bibnamefont {Smelyanskiy}},
  \bibinfo {author} {\bibfnamefont {R.}~\bibnamefont {Babbush}},\ and\ \bibinfo
  {author} {\bibfnamefont {H.}~\bibnamefont {Neven}},\ }\bibfield  {title}
  {\bibinfo {title} {Barren plateaus in quantum neural network training
  landscapes},\ }\href
  {https://doi.org/https://doi.org/10.1038/s41467-018-07090-4} {\bibfield
  {journal} {\bibinfo  {journal} {Nat. Commun.}\ }\textbf {\bibinfo {volume}
  {9}},\ \bibinfo {pages} {1} (\bibinfo {year} {2018})}\BibitemShut {NoStop}%
\bibitem [{\citenamefont {Cerezo}\ \emph
  {et~al.}(2021{\natexlab{a}})\citenamefont {Cerezo}, \citenamefont {Sone},
  \citenamefont {Volkoff}, \citenamefont {Cincio},\ and\ \citenamefont
  {Coles}}]{cerezo2021b}%
  \BibitemOpen
  \bibfield  {author} {\bibinfo {author} {\bibfnamefont {M.}~\bibnamefont
  {Cerezo}}, \bibinfo {author} {\bibfnamefont {A.}~\bibnamefont {Sone}},
  \bibinfo {author} {\bibfnamefont {T.}~\bibnamefont {Volkoff}}, \bibinfo
  {author} {\bibfnamefont {L.}~\bibnamefont {Cincio}},\ and\ \bibinfo {author}
  {\bibfnamefont {P.~J.}\ \bibnamefont {Coles}},\ }\bibfield  {title} {\bibinfo
  {title} {Cost function dependent barren plateaus in shallow parametrized
  quantum circuits},\ }\href {https://doi.org/10.1038/s41467-021-21728-w}
  {\bibfield  {journal} {\bibinfo  {journal} {Nat. Commun.}\ }\textbf {\bibinfo
  {volume} {12}},\ \bibinfo {pages} {1791} (\bibinfo {year}
  {2021}{\natexlab{a}})}\BibitemShut {NoStop}%
\bibitem [{\citenamefont {Sharma}\ \emph {et~al.}(2022)\citenamefont {Sharma},
  \citenamefont {Cerezo}, \citenamefont {Cincio},\ and\ \citenamefont
  {Coles}}]{sharma2022}%
  \BibitemOpen
  \bibfield  {author} {\bibinfo {author} {\bibfnamefont {K.}~\bibnamefont
  {Sharma}}, \bibinfo {author} {\bibfnamefont {M.}~\bibnamefont {Cerezo}},
  \bibinfo {author} {\bibfnamefont {L.}~\bibnamefont {Cincio}},\ and\ \bibinfo
  {author} {\bibfnamefont {P.~J.}\ \bibnamefont {Coles}},\ }\bibfield  {title}
  {\bibinfo {title} {{Trainability of Dissipative Perceptron-Based Quantum
  Neural Networks}},\ }\href {https://doi.org/10.1103/PhysRevLett.128.180505}
  {\bibfield  {journal} {\bibinfo  {journal} {Phys. Rev. Lett.}\ }\textbf
  {\bibinfo {volume} {128}},\ \bibinfo {pages} {180505} (\bibinfo {year}
  {2022})}\BibitemShut {NoStop}%
\bibitem [{\citenamefont {Cerezo}\ \emph {et~al.}(2023)\citenamefont {Cerezo},
  \citenamefont {Larocca}, \citenamefont {Garc{\'\i}a-Mart{\'\i}n},
  \citenamefont {Diaz}, \citenamefont {Braccia}, \citenamefont {Fontana},
  \citenamefont {Rudolph}, \citenamefont {Bermejo}, \citenamefont {Ijaz},
  \citenamefont {Thanasilp} \emph {et~al.}}]{cerezo2023}%
  \BibitemOpen
  \bibfield  {author} {\bibinfo {author} {\bibfnamefont {M.}~\bibnamefont
  {Cerezo}}, \bibinfo {author} {\bibfnamefont {M.}~\bibnamefont {Larocca}},
  \bibinfo {author} {\bibfnamefont {D.}~\bibnamefont
  {Garc{\'\i}a-Mart{\'\i}n}}, \bibinfo {author} {\bibfnamefont {N.~L.}\
  \bibnamefont {Diaz}}, \bibinfo {author} {\bibfnamefont {P.}~\bibnamefont
  {Braccia}}, \bibinfo {author} {\bibfnamefont {E.}~\bibnamefont {Fontana}},
  \bibinfo {author} {\bibfnamefont {M.~S.}\ \bibnamefont {Rudolph}}, \bibinfo
  {author} {\bibfnamefont {P.}~\bibnamefont {Bermejo}}, \bibinfo {author}
  {\bibfnamefont {A.}~\bibnamefont {Ijaz}}, \bibinfo {author} {\bibfnamefont
  {S.}~\bibnamefont {Thanasilp}}, \emph {et~al.},\ }\bibfield  {title}
  {\bibinfo {title} {{Does provable absence of barren plateaus imply classical
  simulability? Or, why we need to rethink variational quantum computing}},\
  }\href {https://doi.org/10.48550/arXiv.2312.09121} {\bibfield  {journal}
  {\bibinfo  {journal} {arXiv:2312.09121}\ } (\bibinfo {year}
  {2023})}\BibitemShut {NoStop}%
\bibitem [{\citenamefont {Larocca}\ \emph {et~al.}(2024)\citenamefont
  {Larocca}, \citenamefont {Thanasilp}, \citenamefont {Wang}, \citenamefont
  {Sharma}, \citenamefont {Biamonte}, \citenamefont {Coles}, \citenamefont
  {Cincio}, \citenamefont {McClean}, \citenamefont {Holmes},\ and\
  \citenamefont {Cerezo}}]{larocca2024}%
  \BibitemOpen
  \bibfield  {author} {\bibinfo {author} {\bibfnamefont {M.}~\bibnamefont
  {Larocca}}, \bibinfo {author} {\bibfnamefont {S.}~\bibnamefont {Thanasilp}},
  \bibinfo {author} {\bibfnamefont {S.}~\bibnamefont {Wang}}, \bibinfo {author}
  {\bibfnamefont {K.}~\bibnamefont {Sharma}}, \bibinfo {author} {\bibfnamefont
  {J.}~\bibnamefont {Biamonte}}, \bibinfo {author} {\bibfnamefont {P.~J.}\
  \bibnamefont {Coles}}, \bibinfo {author} {\bibfnamefont {L.}~\bibnamefont
  {Cincio}}, \bibinfo {author} {\bibfnamefont {J.~R.}\ \bibnamefont {McClean}},
  \bibinfo {author} {\bibfnamefont {Z.}~\bibnamefont {Holmes}},\ and\ \bibinfo
  {author} {\bibfnamefont {M.}~\bibnamefont {Cerezo}},\ }\bibfield  {title}
  {\bibinfo {title} {A review of barren plateaus in variational quantum
  computing},\ }\href {https://doi.org/10.48550/arXiv.2405.00781} {\bibfield
  {journal} {\bibinfo  {journal} {arXiv:2405.00781}\ } (\bibinfo {year}
  {2024})}\BibitemShut {NoStop}%
\bibitem [{\citenamefont {Ragone}\ \emph {et~al.}(2024)\citenamefont {Ragone},
  \citenamefont {Bakalov}, \citenamefont {Sauvage}, \citenamefont {Kemper},
  \citenamefont {Ortiz~Marrero}, \citenamefont {Larocca},\ and\ \citenamefont
  {Cerezo}}]{ragone2024}%
  \BibitemOpen
  \bibfield  {author} {\bibinfo {author} {\bibfnamefont {M.}~\bibnamefont
  {Ragone}}, \bibinfo {author} {\bibfnamefont {B.~N.}\ \bibnamefont {Bakalov}},
  \bibinfo {author} {\bibfnamefont {F.}~\bibnamefont {Sauvage}}, \bibinfo
  {author} {\bibfnamefont {A.~F.}\ \bibnamefont {Kemper}}, \bibinfo {author}
  {\bibfnamefont {C.}~\bibnamefont {Ortiz~Marrero}}, \bibinfo {author}
  {\bibfnamefont {M.}~\bibnamefont {Larocca}},\ and\ \bibinfo {author}
  {\bibfnamefont {M.}~\bibnamefont {Cerezo}},\ }\bibfield  {title} {\bibinfo
  {title} {A lie algebraic theory of barren plateaus for deep parameterized
  quantum circuits},\ }\href {https://doi.org/10.1038/s41467-024-49909-3}
  {\bibfield  {journal} {\bibinfo  {journal} {Nat. Commun.}\ }\textbf {\bibinfo
  {volume} {15}},\ \bibinfo {pages} {7172} (\bibinfo {year}
  {2024})}\BibitemShut {NoStop}%
\bibitem [{\citenamefont {Or{\'u}s}(2014)}]{orus2014}%
  \BibitemOpen
  \bibfield  {author} {\bibinfo {author} {\bibfnamefont {R.}~\bibnamefont
  {Or{\'u}s}},\ }\bibfield  {title} {\bibinfo {title} {{A practical
  introduction to tensor networks: Matrix product states and projected
  entangled pair states}},\ }\href {https://doi.org/10.1016/j.aop.2014.06.013}
  {\bibfield  {journal} {\bibinfo  {journal} {Ann. Phys.}\ }\textbf {\bibinfo
  {volume} {349}},\ \bibinfo {pages} {117} (\bibinfo {year}
  {2014})}\BibitemShut {NoStop}%
\bibitem [{\citenamefont {Paeckel}\ \emph {et~al.}(2019)\citenamefont
  {Paeckel}, \citenamefont {K{\"o}hler}, \citenamefont {Swoboda}, \citenamefont
  {Manmana}, \citenamefont {Schollw{\"o}ck},\ and\ \citenamefont
  {Hubig}}]{paeckel2019}%
  \BibitemOpen
  \bibfield  {author} {\bibinfo {author} {\bibfnamefont {S.}~\bibnamefont
  {Paeckel}}, \bibinfo {author} {\bibfnamefont {T.}~\bibnamefont {K{\"o}hler}},
  \bibinfo {author} {\bibfnamefont {A.}~\bibnamefont {Swoboda}}, \bibinfo
  {author} {\bibfnamefont {S.~R.}\ \bibnamefont {Manmana}}, \bibinfo {author}
  {\bibfnamefont {U.}~\bibnamefont {Schollw{\"o}ck}},\ and\ \bibinfo {author}
  {\bibfnamefont {C.}~\bibnamefont {Hubig}},\ }\bibfield  {title} {\bibinfo
  {title} {Time-evolution methods for matrix-product states},\ }\href
  {https://doi.org/10.1016/j.aop.2019.167998} {\bibfield  {journal} {\bibinfo
  {journal} {Ann. Phys.}\ }\textbf {\bibinfo {volume} {411}},\ \bibinfo {pages}
  {167998} (\bibinfo {year} {2019})}\BibitemShut {NoStop}%
\bibitem [{\citenamefont {Lorenzo}\ \emph {et~al.}(2017)\citenamefont
  {Lorenzo}, \citenamefont {Ciccarello},\ and\ \citenamefont
  {Palma}}]{lorenzo2017}%
  \BibitemOpen
  \bibfield  {author} {\bibinfo {author} {\bibfnamefont {S.}~\bibnamefont
  {Lorenzo}}, \bibinfo {author} {\bibfnamefont {F.}~\bibnamefont
  {Ciccarello}},\ and\ \bibinfo {author} {\bibfnamefont {G.~M.}\ \bibnamefont
  {Palma}},\ }\bibfield  {title} {\bibinfo {title} {Composite quantum collision
  models},\ }\href {https://doi.org/10.1103/PhysRevA.96.032107} {\bibfield
  {journal} {\bibinfo  {journal} {Phys. Rev. A}\ }\textbf {\bibinfo {volume}
  {96}},\ \bibinfo {pages} {032107} (\bibinfo {year} {2017})}\BibitemShut
  {NoStop}%
\bibitem [{\citenamefont {Ciccarello}(2017)}]{ciccarello2017}%
  \BibitemOpen
  \bibfield  {author} {\bibinfo {author} {\bibfnamefont {F.}~\bibnamefont
  {Ciccarello}},\ }\bibfield  {title} {\bibinfo {title} {Collision models in
  quantum optics},\ }\href {https://doi.org/10.1515/qmetro-2017-0007}
  {\bibfield  {journal} {\bibinfo  {journal} {Quantum Meas. Quantum Metrol.}\
  }\textbf {\bibinfo {volume} {4}},\ \bibinfo {pages} {53} (\bibinfo {year}
  {2017})}\BibitemShut {NoStop}%
\bibitem [{\citenamefont {Ciccarello}\ \emph {et~al.}(2022)\citenamefont
  {Ciccarello}, \citenamefont {Lorenzo}, \citenamefont {Giovannetti},\ and\
  \citenamefont {Palma}}]{ciccarello2022}%
  \BibitemOpen
  \bibfield  {author} {\bibinfo {author} {\bibfnamefont {F.}~\bibnamefont
  {Ciccarello}}, \bibinfo {author} {\bibfnamefont {S.}~\bibnamefont {Lorenzo}},
  \bibinfo {author} {\bibfnamefont {V.}~\bibnamefont {Giovannetti}},\ and\
  \bibinfo {author} {\bibfnamefont {G.~M.}\ \bibnamefont {Palma}},\ }\bibfield
  {title} {\bibinfo {title} {{Quantum collision models: Open system dynamics
  from repeated interactions}},\ }\href
  {https://doi.org/https://doi.org/10.1016/j.physrep.2022.01.001} {\bibfield
  {journal} {\bibinfo  {journal} {Phys. Rep.}\ }\textbf {\bibinfo {volume}
  {954}},\ \bibinfo {pages} {1} (\bibinfo {year} {2022})}\BibitemShut {NoStop}%
\bibitem [{\citenamefont {Cattaneo}\ \emph {et~al.}(2021)\citenamefont
  {Cattaneo}, \citenamefont {De~Chiara}, \citenamefont {Maniscalco},
  \citenamefont {Zambrini},\ and\ \citenamefont {Giorgi}}]{cattaneo2021}%
  \BibitemOpen
  \bibfield  {author} {\bibinfo {author} {\bibfnamefont {M.}~\bibnamefont
  {Cattaneo}}, \bibinfo {author} {\bibfnamefont {G.}~\bibnamefont {De~Chiara}},
  \bibinfo {author} {\bibfnamefont {S.}~\bibnamefont {Maniscalco}}, \bibinfo
  {author} {\bibfnamefont {R.}~\bibnamefont {Zambrini}},\ and\ \bibinfo
  {author} {\bibfnamefont {G.~L.}\ \bibnamefont {Giorgi}},\ }\bibfield  {title}
  {\bibinfo {title} {{Collision Models Can Efficiently Simulate Any
  Multipartite Markovian Quantum Dynamics}},\ }\href
  {https://doi.org/10.1103/PhysRevLett.126.130403} {\bibfield  {journal}
  {\bibinfo  {journal} {Phys. Rev. Lett.}\ }\textbf {\bibinfo {volume} {126}},\
  \bibinfo {pages} {130403} (\bibinfo {year} {2021})}\BibitemShut {NoStop}%
\bibitem [{\citenamefont {Cattaneo}\ \emph {et~al.}(2022)\citenamefont
  {Cattaneo}, \citenamefont {Giorgi}, \citenamefont {Zambrini},\ and\
  \citenamefont {Maniscalco}}]{cattaneo2022}%
  \BibitemOpen
  \bibfield  {author} {\bibinfo {author} {\bibfnamefont {M.}~\bibnamefont
  {Cattaneo}}, \bibinfo {author} {\bibfnamefont {G.~L.}\ \bibnamefont
  {Giorgi}}, \bibinfo {author} {\bibfnamefont {R.}~\bibnamefont {Zambrini}},\
  and\ \bibinfo {author} {\bibfnamefont {S.}~\bibnamefont {Maniscalco}},\
  }\bibfield  {title} {\bibinfo {title} {{A brief journey through collision
  models for multipartite open quantum dynamics}},\ }\href
  {https://doi.org/https://doi.org/10.1142/S1230161222500159} {\bibfield
  {journal} {\bibinfo  {journal} {Open Syst. Inf. Dyn.}\ }\textbf {\bibinfo
  {volume} {29}},\ \bibinfo {pages} {2250015} (\bibinfo {year}
  {2022})}\BibitemShut {NoStop}%
\bibitem [{\citenamefont {Lindblad}(1976)}]{lindblad1976}%
  \BibitemOpen
  \bibfield  {author} {\bibinfo {author} {\bibfnamefont {G.}~\bibnamefont
  {Lindblad}},\ }\bibfield  {title} {\bibinfo {title} {On the generators of
  quantum dynamical semigroups},\ }\href
  {https://doi.org/https://doi.org/10.1007/BF01608499} {\bibfield  {journal}
  {\bibinfo  {journal} {Commun. Math. Phys.}\ }\textbf {\bibinfo {volume}
  {48}},\ \bibinfo {pages} {119} (\bibinfo {year} {1976})}\BibitemShut
  {NoStop}%
\bibitem [{\citenamefont {Gorini}\ \emph {et~al.}(1976)\citenamefont {Gorini},
  \citenamefont {Kossakowski},\ and\ \citenamefont {Sudarshan}}]{gorini1976}%
  \BibitemOpen
  \bibfield  {author} {\bibinfo {author} {\bibfnamefont {V.}~\bibnamefont
  {Gorini}}, \bibinfo {author} {\bibfnamefont {A.}~\bibnamefont
  {Kossakowski}},\ and\ \bibinfo {author} {\bibfnamefont {E.~C.~G.}\
  \bibnamefont {Sudarshan}},\ }\bibfield  {title} {\bibinfo {title}
  {{Completely positive dynamical semigroups of N-level systems}},\ }\href
  {https://doi.org/https://doi.org/10.1063/1.522979} {\bibfield  {journal}
  {\bibinfo  {journal} {J. Math. Phys.}\ }\textbf {\bibinfo {volume} {17}},\
  \bibinfo {pages} {821} (\bibinfo {year} {1976})}\BibitemShut {NoStop}%
\bibitem [{\citenamefont {Minganti}\ \emph {et~al.}(2018)\citenamefont
  {Minganti}, \citenamefont {Biella}, \citenamefont {Bartolo},\ and\
  \citenamefont {Ciuti}}]{minganti2018}%
  \BibitemOpen
  \bibfield  {author} {\bibinfo {author} {\bibfnamefont {F.}~\bibnamefont
  {Minganti}}, \bibinfo {author} {\bibfnamefont {A.}~\bibnamefont {Biella}},
  \bibinfo {author} {\bibfnamefont {N.}~\bibnamefont {Bartolo}},\ and\ \bibinfo
  {author} {\bibfnamefont {C.}~\bibnamefont {Ciuti}},\ }\bibfield  {title}
  {\bibinfo {title} {Spectral theory of liouvillians for dissipative phase
  transitions},\ }\href {https://doi.org/10.1103/PhysRevA.98.042118} {\bibfield
   {journal} {\bibinfo  {journal} {Phys. Rev. A}\ }\textbf {\bibinfo {volume}
  {98}},\ \bibinfo {pages} {042118} (\bibinfo {year} {2018})}\BibitemShut
  {NoStop}%
\bibitem [{SM()}]{SM}%
  \BibitemOpen
  \href@noop {} {\bibinfo  {journal} {See Supplemental Material at [URL will be
  inserted by publisher] for the details about the calculations and the
  tensor-network simulations.}\ }\BibitemShut {NoStop}%
\bibitem [{\citenamefont {Navez}\ and\ \citenamefont
  {Sch\"utzhold}(2010)}]{PRA_BBGKY}%
  \BibitemOpen
\bibfield  {journal} {  }\bibfield  {author} {\bibinfo {author} {\bibfnamefont
  {P.}~\bibnamefont {Navez}}\ and\ \bibinfo {author} {\bibfnamefont
  {R.}~\bibnamefont {Sch\"utzhold}},\ }\bibfield  {title} {\bibinfo {title}
  {Emergence of coherence in the mott-insulator--superfluid quench of the
  bose-hubbard model},\ }\href {https://doi.org/10.1103/PhysRevA.82.063603}
  {\bibfield  {journal} {\bibinfo  {journal} {Phys. Rev. A}\ }\textbf {\bibinfo
  {volume} {82}},\ \bibinfo {pages} {063603} (\bibinfo {year}
  {2010})}\BibitemShut {NoStop}%
\bibitem [{\citenamefont {Bisio}\ \emph {et~al.}(2010)\citenamefont {Bisio},
  \citenamefont {Chiribella}, \citenamefont {D'Ariano}, \citenamefont
  {Facchini},\ and\ \citenamefont {Perinotti}}]{bisio2010}%
  \BibitemOpen
  \bibfield  {author} {\bibinfo {author} {\bibfnamefont {A.}~\bibnamefont
  {Bisio}}, \bibinfo {author} {\bibfnamefont {G.}~\bibnamefont {Chiribella}},
  \bibinfo {author} {\bibfnamefont {G.~M.}\ \bibnamefont {D'Ariano}}, \bibinfo
  {author} {\bibfnamefont {S.}~\bibnamefont {Facchini}},\ and\ \bibinfo
  {author} {\bibfnamefont {P.}~\bibnamefont {Perinotti}},\ }\bibfield  {title}
  {\bibinfo {title} {Optimal quantum learning of a unitary transformation},\
  }\href {https://link.aps.org/doi/10.1103/PhysRevA.81.032324} {\bibfield
  {journal} {\bibinfo  {journal} {Phys. Rev. A}\ }\textbf {\bibinfo {volume}
  {81}},\ \bibinfo {pages} {032324} (\bibinfo {year} {2010})}\BibitemShut
  {NoStop}%
\bibitem [{\citenamefont {Cincio}\ \emph {et~al.}(2018)\citenamefont {Cincio},
  \citenamefont {Subaşı}, \citenamefont {Sornborger},\ and\ \citenamefont
  {Coles}}]{cincio2018}%
  \BibitemOpen
  \bibfield  {author} {\bibinfo {author} {\bibfnamefont {L.}~\bibnamefont
  {Cincio}}, \bibinfo {author} {\bibfnamefont {Y.}~\bibnamefont {Subaşı}},
  \bibinfo {author} {\bibfnamefont {A.~T.}\ \bibnamefont {Sornborger}},\ and\
  \bibinfo {author} {\bibfnamefont {P.~J.}\ \bibnamefont {Coles}},\ }\bibfield
  {title} {\bibinfo {title} {Learning the quantum algorithm for state
  overlap},\ }\href {https://dx.doi.org/10.1088/1367-2630/aae94a} {\bibfield
  {journal} {\bibinfo  {journal} {New J. Phys.}\ }\textbf {\bibinfo {volume}
  {20}},\ \bibinfo {pages} {113022} (\bibinfo {year} {2018})}\BibitemShut
  {NoStop}%
\bibitem [{\citenamefont {Cerezo}\ \emph {et~al.}(2020)\citenamefont {Cerezo},
  \citenamefont {Poremba}, \citenamefont {Cincio},\ and\ \citenamefont
  {Coles}}]{cerezo2020}%
  \BibitemOpen
  \bibfield  {author} {\bibinfo {author} {\bibfnamefont {M.}~\bibnamefont
  {Cerezo}}, \bibinfo {author} {\bibfnamefont {A.}~\bibnamefont {Poremba}},
  \bibinfo {author} {\bibfnamefont {L.}~\bibnamefont {Cincio}},\ and\ \bibinfo
  {author} {\bibfnamefont {P.~J.}\ \bibnamefont {Coles}},\ }\bibfield  {title}
  {\bibinfo {title} {Variational {Q}uantum {F}idelity {E}stimation},\ }\href
  {https://doi.org/10.22331/q-2020-03-26-248} {\bibfield  {journal} {\bibinfo
  {journal} {{Quantum}}\ }\textbf {\bibinfo {volume} {4}},\ \bibinfo {pages}
  {248} (\bibinfo {year} {2020})}\BibitemShut {NoStop}%
\bibitem [{\citenamefont {Xiao}\ \emph {et~al.}(2022)\citenamefont {Xiao},
  \citenamefont {Chen},\ and\ \citenamefont {Xu}}]{xiao2022}%
  \BibitemOpen
  \bibfield  {author} {\bibinfo {author} {\bibfnamefont {H.}~\bibnamefont
  {Xiao}}, \bibinfo {author} {\bibfnamefont {X.}~\bibnamefont {Chen}},\ and\
  \bibinfo {author} {\bibfnamefont {J.}~\bibnamefont {Xu}},\ }\bibfield
  {title} {\bibinfo {title} {Using a deep quantum neural network to enhance the
  fidelity of quantum convolutional codes},\ }\href
  {https://www.mdpi.com/2076-3417/12/11/5662} {\bibfield  {journal} {\bibinfo
  {journal} {Appl. Sci.}\ }\textbf {\bibinfo {volume} {12}} (\bibinfo {year}
  {2022})}\BibitemShut {NoStop}%
\bibitem [{\citenamefont {Wu}\ \emph {et~al.}(2024)\citenamefont {Wu},
  \citenamefont {Fu}, \citenamefont {Zhu}, \citenamefont {Zhang}, \citenamefont
  {Xie},\ and\ \citenamefont {Li}}]{wu2024}%
  \BibitemOpen
  \bibfield  {author} {\bibinfo {author} {\bibfnamefont {J.}~\bibnamefont
  {Wu}}, \bibinfo {author} {\bibfnamefont {H.}~\bibnamefont {Fu}}, \bibinfo
  {author} {\bibfnamefont {M.}~\bibnamefont {Zhu}}, \bibinfo {author}
  {\bibfnamefont {H.}~\bibnamefont {Zhang}}, \bibinfo {author} {\bibfnamefont
  {W.}~\bibnamefont {Xie}},\ and\ \bibinfo {author} {\bibfnamefont {X.-Y.}\
  \bibnamefont {Li}},\ }\bibfield  {title} {\bibinfo {title} {Quantum circuit
  autoencoder},\ }\href {https://doi.org/10.1103/PhysRevA.109.032623}
  {\bibfield  {journal} {\bibinfo  {journal} {Phys. Rev. A}\ }\textbf {\bibinfo
  {volume} {109}},\ \bibinfo {pages} {032623} (\bibinfo {year}
  {2024})}\BibitemShut {NoStop}%
\bibitem [{\citenamefont {Ewald}(2025)}]{ewald2025}%
  \BibitemOpen
  \bibfield  {author} {\bibinfo {author} {\bibfnamefont {D.}~\bibnamefont
  {Ewald}},\ }\bibfield  {title} {\bibinfo {title} {The proposal of a fully
  quantum neural network and fidelity-driven training using directional
  gradients for multi-class classification},\ }\href
  {https://www.mdpi.com/2079-9292/14/11/2189} {\bibfield  {journal} {\bibinfo
  {journal} {Electronics}\ }\textbf {\bibinfo {volume} {14}} (\bibinfo {year}
  {2025})}\BibitemShut {NoStop}%
\bibitem [{\citenamefont {Anderson}(1967)}]{anderson1967}%
  \BibitemOpen
  \bibfield  {author} {\bibinfo {author} {\bibfnamefont {P.~W.}\ \bibnamefont
  {Anderson}},\ }\bibfield  {title} {\bibinfo {title} {Infrared catastrophe in
  fermi gases with local scattering potentials},\ }\href
  {https://doi.org/10.1103/PhysRevLett.18.1049} {\bibfield  {journal} {\bibinfo
   {journal} {Phys. Rev. Lett.}\ }\textbf {\bibinfo {volume} {18}},\ \bibinfo
  {pages} {1049} (\bibinfo {year} {1967})}\BibitemShut {NoStop}%
\bibitem [{\citenamefont {Cerezo}\ \emph
  {et~al.}(2021{\natexlab{b}})\citenamefont {Cerezo}, \citenamefont
  {Arrasmith}, \citenamefont {Babbush}, \citenamefont {Benjamin}, \citenamefont
  {Endo}, \citenamefont {Fujii}, \citenamefont {McClean}, \citenamefont
  {Mitarai}, \citenamefont {Yuan}, \citenamefont {Cincio} \emph
  {et~al.}}]{cerezo2021a}%
  \BibitemOpen
  \bibfield  {author} {\bibinfo {author} {\bibfnamefont {M.}~\bibnamefont
  {Cerezo}}, \bibinfo {author} {\bibfnamefont {A.}~\bibnamefont {Arrasmith}},
  \bibinfo {author} {\bibfnamefont {R.}~\bibnamefont {Babbush}}, \bibinfo
  {author} {\bibfnamefont {S.~C.}\ \bibnamefont {Benjamin}}, \bibinfo {author}
  {\bibfnamefont {S.}~\bibnamefont {Endo}}, \bibinfo {author} {\bibfnamefont
  {K.}~\bibnamefont {Fujii}}, \bibinfo {author} {\bibfnamefont {J.~R.}\
  \bibnamefont {McClean}}, \bibinfo {author} {\bibfnamefont {K.}~\bibnamefont
  {Mitarai}}, \bibinfo {author} {\bibfnamefont {X.}~\bibnamefont {Yuan}},
  \bibinfo {author} {\bibfnamefont {L.}~\bibnamefont {Cincio}}, \emph
  {et~al.},\ }\bibfield  {title} {\bibinfo {title} {Variational quantum
  algorithms},\ }\href {https://doi.org/10.1038/s42254-021-00348-9} {\bibfield
  {journal} {\bibinfo  {journal} {Nat. Rev. Phys.}\ }\textbf {\bibinfo {volume}
  {3}},\ \bibinfo {pages} {625} (\bibinfo {year}
  {2021}{\natexlab{b}})}\BibitemShut {NoStop}%
\bibitem [{\citenamefont {Potts}(1952)}]{potts1952}%
  \BibitemOpen
  \bibfield  {author} {\bibinfo {author} {\bibfnamefont {R.~B.}\ \bibnamefont
  {Potts}},\ }\bibfield  {title} {\bibinfo {title} {Some generalized
  order-disorder transformations},\ }\href
  {https://doi.org/10.1017/S0305004100027419} {\bibfield  {journal} {\bibinfo
  {journal} {Math. Proc. Camb. Philos. Soc.}\ }\textbf {\bibinfo {volume}
  {48}},\ \bibinfo {pages} {106–109} (\bibinfo {year} {1952})}\BibitemShut
  {NoStop}%
\bibitem [{\citenamefont {S\'olyom}(1981)}]{solyom1981a}%
  \BibitemOpen
  \bibfield  {author} {\bibinfo {author} {\bibfnamefont {J.}~\bibnamefont
  {S\'olyom}},\ }\bibfield  {title} {\bibinfo {title} {Duality of the block
  transformation and decimation for quantum spin systems},\ }\href
  {https://doi.org/10.1103/PhysRevB.24.230} {\bibfield  {journal} {\bibinfo
  {journal} {Phys. Rev. B}\ }\textbf {\bibinfo {volume} {24}},\ \bibinfo
  {pages} {230} (\bibinfo {year} {1981})}\BibitemShut {NoStop}%
\bibitem [{\citenamefont {S\'olyom}\ and\ \citenamefont
  {Pfeuty}(1981)}]{solyom1981b}%
  \BibitemOpen
  \bibfield  {author} {\bibinfo {author} {\bibfnamefont {J.}~\bibnamefont
  {S\'olyom}}\ and\ \bibinfo {author} {\bibfnamefont {P.}~\bibnamefont
  {Pfeuty}},\ }\bibfield  {title} {\bibinfo {title} {Renormalization-group
  study of the hamiltonian version of the potts model},\ }\href
  {https://doi.org/10.1103/PhysRevB.24.218} {\bibfield  {journal} {\bibinfo
  {journal} {Phys. Rev. B}\ }\textbf {\bibinfo {volume} {24}},\ \bibinfo
  {pages} {218} (\bibinfo {year} {1981})}\BibitemShut {NoStop}%
\bibitem [{\citenamefont {Ptaszy\ifmmode~\acute{n}\else \'{n}\fi{}ski}\ and\
  \citenamefont {Esposito}(2024)}]{esposito2024}%
  \BibitemOpen
  \bibfield  {author} {\bibinfo {author} {\bibfnamefont {K.}~\bibnamefont
  {Ptaszy\ifmmode~\acute{n}\else \'{n}\fi{}ski}}\ and\ \bibinfo {author}
  {\bibfnamefont {M.}~\bibnamefont {Esposito}},\ }\bibfield  {title} {\bibinfo
  {title} {Dynamical signatures of discontinuous phase transitions: How phase
  coexistence determines exponential versus power-law scaling},\ }\href
  {https://doi.org/10.1103/PhysRevE.110.044134} {\bibfield  {journal} {\bibinfo
   {journal} {Phys. Rev. E}\ }\textbf {\bibinfo {volume} {110}},\ \bibinfo
  {pages} {044134} (\bibinfo {year} {2024})}\BibitemShut {NoStop}%
\bibitem [{\citenamefont {Causer}\ \emph {et~al.}(2025)\citenamefont {Causer},
  \citenamefont {Ba\~nuls},\ and\ \citenamefont {Garrahan}}]{causer2025}%
  \BibitemOpen
  \bibfield  {author} {\bibinfo {author} {\bibfnamefont {L.}~\bibnamefont
  {Causer}}, \bibinfo {author} {\bibfnamefont {M.~C.}\ \bibnamefont
  {Ba\~nuls}},\ and\ \bibinfo {author} {\bibfnamefont {J.~P.}\ \bibnamefont
  {Garrahan}},\ }\bibfield  {title} {\bibinfo {title} {Dynamical heterogeneity
  and large deviations in the open quantum east glass model from tensor
  networks},\ }\href {https://doi.org/10.1103/PhysRevB.111.L060303} {\bibfield
  {journal} {\bibinfo  {journal} {Phys. Rev. B}\ }\textbf {\bibinfo {volume}
  {111}},\ \bibinfo {pages} {L060303} (\bibinfo {year} {2025})}\BibitemShut
  {NoStop}%
\end{thebibliography}%

\setcounter{equation}{0}
\setcounter{figure}{0}
\setcounter{table}{0}
\renewcommand{\theequation}{S\arabic{equation}}
\renewcommand{\thefigure}{S\arabic{figure}}

\makeatletter
\renewcommand{\theequation}{S\arabic{figure}}
\renewcommand{\thefigure}{S\arabic{figure}}

\onecolumngrid
\newpage

\setcounter{page}{1}

\setcounter{secnumdepth}{3}
\pagestyle{plain}

\begin{center}
{\Large SUPPLEMENTAL MATERIAL}
\setcounter{page}{1}
\end{center}
\begin{center}
\vspace{0.8cm}
{\Large Training the classification capability of large-scale quantum cellular automata}
\end{center}
\begin{center}
Mario Boneberg,$^{1}$, Simon Kochsiek$^{1}$, Gabriele Perfetto$^{1}$ and Igor Lesanovsky$^{1,2}$
\end{center}
\begin{center}
$^1$ {\em Institut f\"ur Theoretische Physik, Universit\"at Tübingen and Center for Integrated Quantum Science and Technology,}\\
{\em  Auf der Morgenstelle 14, 72076 T\"ubingen, Germany}\\
$^2$ {\em School of Physics and Astronomy and Centre for the Mathematics}\\
{\em  and Theoretical Physics of Quantum Non-Equilibrium Systems,}\\
{\em  The University of Nottingham, Nottingham, NG7 2RD, United Kingdom}
\end{center}

\setcounter{equation}{0}
\setcounter{figure}{0}
\setcounter{table}{0}
\setcounter{page}{1}
\makeatletter
\renewcommand{\theequation}{S\arabic{equation}}
\renewcommand{\thefigure}{S\arabic{figure}}

\makeatletter
\renewcommand{\theequation}{S\arabic{equation}}
\renewcommand{\thefigure}{S\arabic{figure}}

\renewcommand{\bibnumfmt}[1]{[S#1]}
\renewcommand{\citenumfont}[1]{S#1}

\onecolumngrid

\setcounter{secnumdepth}{3}


\section{Derivation of the equations of motion including nearest neighbor correlations}
In this section, we detail the derivation of the equations of motion in Eq.~(7) of the main text. Therein the inclusion of spatial correlations between nearest neighbor sites leads to a set of twelve differential equations \eqref{app:correlations_eom_final} below. These equations have been used to produce the phase diagram in Fig.~2(b) of the main text. When spatial correlations are neglected, the equations \eqref{app:correlations_eom_final} reduce to the smaller set \eqref{app:mean_field_eom_final} given by the three mean-field equations for the magnetization in directions $x$, $y$ and $z$, respectively. The mean-field equations \eqref{app:mean_field_eom_final} have been used to produce the phase diagram in Fig.~2(a) of the main text. In this section, we also use the following notation for the one ($\rho_a$) and two-sites ($\rho_{ab}$) reduced density matrices
\begin{equation}
\rho_{a}=\mbox{Tr}_{\neq a}[\rho], \quad \mbox{and} \quad \rho_{ab}=\mbox{Tr}_{\neq ab}[\rho], 
\label{app_partial_trace}  
\end{equation}
obtained after a partial traces of the full density matrix $\rho$. Notice that differently from the main text, the lattice site indices $a,b$ are reported as subscripts, and, for brevity of notation, we omit the time dependency. For the one-body density matrix the partial trace $\mbox{Tr}_{\neq a}[\dots]$ is taken over all sites but $a$, while for the two-body reduced state on all sites but the pair $a,b$ as $\mbox{Tr}_{\neq a,b}[\dots]$. We consider, for the sake of generality, the system to be defined on an hypercubic lattice with $N$ lattice sites in $d$ spatial dimensions. In Fig.~(2), of the main text, we specialized to the one-dimensional case.

We start by writing the trial state $\rho_c$ for the density matrix which includes nearest neighbour correlations as 
\begin{equation}
\rho_c= \rho_{\mathrm{MF}} +\sum_{\langle{j,k}\rangle} C_{j,k}\left(\prod_{l \neq j,k}\rho_l \right) \quad \mbox{(trial state including nearest neighbor correlations)}.
\label{app:correlations_trial}
\end{equation}
Here, we defined the mean-field state $\rho_{\mathrm{MF}}$ and nearest neighbor connected correlation function $C_{j,k}$ respectively as
\begin{equation}
\rho_{\mathrm{MF}}=\prod_{j=1}^{N}\rho_j, \quad \mbox{and}\quad C_{j,k}=\rho_{jk}-\rho_j\rho_k.
\label{app:density_matrix_correlations_expression}
\end{equation}
Here the sum $\sum_{\langle{j,k}\rangle}$ runs over pairs of nearest neighbour sites $j$ and $k$. 
The trial state $\rho_c$ thereby exactly includes nearest neighbor correlations of a single bond $(j,k)$ via the connected correlation function $C_{j,k}$ in \eqref{app:density_matrix_correlations_expression}. These correlations are generated in time due to the presence of nearest neighbor coupling proportional to $V$ in the Hamiltonian (see Eq.~(4) of the main text). The Jump operator $J_k$ in Eqs.~(2)-(4) of the main text is, instead, purely a single-body operator. The two-body density matrix $\rho_{jk}$ describing the reduced state on the bond $(j,k)$ is therefore in Eq.~\eqref{app:correlations_trial} \textit{not} factorized in terms of the single-site reduced states $\rho_j$ and $\rho_k$, as it is the case in mean-field. In the latter, indeed, the connected correlation function $C_{j,k}\equiv 0$ vanishes and only the state $\rho_{\mathrm{MF}}$ factorized in terms of the single-site reduced density matrix remains
\begin{equation}
\rho_{\mathrm{MF}}=\prod_{j=1}^N\rho_j \quad \mbox{(mean-field trial state)}.
\label{app:mean_field_state}
\end{equation}

We now therefore derive the equations of motion for the two-body nearest neighbor correlations describing the state \eqref{app:correlations_trial}. To this aim, we conveniently rewrite the correlated trial state \eqref{app:correlations_trial}  as
\begin{align}
\rho_c&=\rho_{\mathrm{MF}} +\sum_{\langle j,k\rangle}(\rho_{jk}-\rho_{j}\rho_{k})\left(\prod_{l\neq j,k} \rho_l \right)=\rho_{\mathrm{MF}}+\sum_{\langle j,j\rangle}\rho_{jk}\left(\prod_{l\neq j,k} \rho_l\right) - \sum_{\langle j,k\rangle} \rho_{j}\rho_k \left(\prod_{l\neq j,k} \rho_l\right) \nonumber \\
&=\rho_{\mathrm{MF}}+\sum_{\langle j,k\rangle}\rho_{jk}\left(\prod_{l\neq j,k} \rho_l\right) - \frac{q N}{2}\rho_{\mathrm{MF}}=\rho_{\mathrm{MF}}\left(1-\frac{qN}{2}\right)+\sum_{\langle j,k\rangle}\rho_{jk}\left(\prod_{l\neq j,k} \rho_l\right),
\label{app:intermediate}
\end{align}
with $q$ denoting the coordination number, i.e., the number of nearest neighbors to a given lattice site. For an hypercubic lattice in $d$ spatial dimensions, one has $q=2d$. The total number of bonds on the lattice is accordingly given by $q N/2$ (the factor $2$ avoids double counting since each bond is shared by two lattice sites). The equation of motion for the reduced state $\rho_{ab}^c=\mbox{Tr}_{\neq ab}[\rho_c]$ over the nearest neighbor bond $\langle a,b\rangle$ is obtained by taking the time derivative of the partial trace of Eqs.~\eqref{app:intermediate} according to the Lindblad equation in Eq.~(2) of the main text. We then obtain
\begin{align}
\frac{\mbox{d}\rho^c_{ab}}{\mbox{d}t} &= \left(1-\frac{q N}{2}\right)\frac{\mbox{d} \,\mbox{Tr}_{\neq ab}[\rho_{\mathrm{MF}}]}{\mbox{d} t} + \frac{\mbox{d}\, \mbox{Tr}_{\neq ab}[\rho_{ab} \prod_{l\neq a,b}\rho_l] }{\mbox{d} t}+\sum_{\langle j,k\rangle \neq \langle a,b\rangle}\frac{\mbox{d}\, \mbox{Tr}_{\neq ab}[\rho_{jk} \prod_{l\neq j,k}\rho_l] }{\mbox{d} t}\nonumber \\
&=\left(1-\frac{q N}{2}\right)\left(\frac{\mbox{d}\rho_a}{\mbox{d}t}\rho_b + \frac{\mbox{d}\rho_b}{\mbox{d}t}\rho_a \right)+\frac{\mbox{d}\, \mbox{Tr}_{\neq ab}[\rho_{ab} \prod_{l\neq a,b}\rho_l] }{\mbox{d} t} +\left(\frac{qN}{2}-1\right)\left(\frac{\mbox{d}\rho_a}{\mbox{d}t}\rho_b + \frac{\mbox{d}\rho_b}{\mbox{d}t}\rho_a \right) \nonumber \\
&=\frac{\mbox{d}\, \mbox{Tr}_{\neq ab}[\rho_{ab} \prod_{l\neq a,b}\rho_l] }{\mbox{d} t}=\frac{\mbox{d} \rho_{ab}}{\mbox{d} t}.
\label{app:correlations_step_1}
\end{align}
In passing from the first to the second line we used translation invariance, so the $qN/2 -1$ terms in the $\sum_{\langle j,k \rangle \neq \langle a,b \rangle}[\dots]$ equally contribute. From the last equality in Eq.~\eqref{app:correlations_step_1} it is clear that the reduced dynamics of the state \eqref{app:density_matrix_correlations_expression} is  obtained by treating exactly the correlations on the bond $\langle a,b\rangle$, while correlations with the surrounding sites are still treated in mean-field way. In order to further simplify the expression in Eq.~\eqref{app:correlations_step_1}, we need to consider a specific structure of the Lindbladian dictating the time evolution. Focusing on the dissipative dynamics in Eqs.~(2)-(4) of the main text, it is then useful to distinguish one-body and two-body terms in the Lindbladian. The former are given by (cf. Eq.~(4) of the main text) 
\begin{equation}
H^{(1)}_j= \frac{\Omega}{2}\sigma^z_j, \quad \mbox{and} \quad J_j=\sqrt{\kappa}\sigma^{-}_{j},
\label{eq:one_body}    
\end{equation}
by the transverse field (proportional to $\Omega$) and by the dissipation (proportional to $V$), while the latter 
\begin{equation}
H_{\langle jk \rangle}^{(2)}=-\frac{V}{4} \sigma^x_{j}\sigma^x_{k}.
\label{eq:two_body}    
\end{equation}
by the ferromagnetic coupling (proportional to $V$). The dynamics of the reduced state in Eq.~\eqref{app:correlations_step_1} for the Lindblad dynamics identified by Eqs.~\eqref{eq:one_body} and \eqref{eq:two_body} is 
\begin{align}
\frac{\mbox{d}\rho^c_{ab}}{\mbox{d}t}=\frac{\mbox{d} \rho_{ab}}{\mbox{d} t}=&-i[H^{(1)}_a,\rho_{ab}]-i[H^{(1)}_b, \rho_{ab}]-i[H^{(2)}_{\langle ab \rangle},\rho_{ab}]+\sum_{j=a,b} \mathcal{D}_j(\rho_{ab}) \nonumber \\
&-i \sum_{l\in nn(b)\neq a}\mbox{Tr}_l\left\{ \right [H_{\langle lb \rangle}^{(2)},\rho_{abl}]  \}-i \sum_{l\in nn(a)\neq b}\mbox{Tr}_l\left\{ \right [H_{\langle la \rangle}^{(2)},\rho_{abl}]  \}
\label{app:hierarchy}
\end{align}
This equation represents an example of the Bogoliubov-Born-Green-
Kirkwood-Yvon hierarchy describing the dynamics of interacting many-body systems, where the equation for the $n$-body reduced density matrix is coupled to the $n+1$-body density matrix, see, e.g., Ref.~\cite{PRA_BBGKY}. In the specific case of \eqref{app:hierarchy}, we see that the dynamics of the two-body density matrix $\rho_{ab}$ is coupled to the three-body density matrix $\rho_{abl}$, which renders \eqref{app:hierarchy} not a closed equation and therefore of no practical utility. Within the set of trial states \eqref{app:correlations_trial}, however, we only consider correlations on a single-bond exactly, while the remainder of the state takes still the mean-field factorized form, as can be seen in \eqref{app:correlations_step_1}. We can therefore proceed further by writing $\rho_{abl}=\rho_{ab}\rho_l$ in \eqref{app:hierarchy}. In this way, one obtains
\begin{align}
\frac{\mbox{d} \rho_{ab}}{\mbox{d} t}
=&-i[H^{(1)}_a,\rho_{ab}]-i[H^{(1)}_b, \rho_{ab}]+\sum_{j=a,b} \mathcal{D}_j(\rho_{ab}) \nonumber \\
&-i[H^{(2)}_{\langle ab\rangle},\rho_{ab}]-i \sum_{l\in nn(b)\neq a}\mbox{Tr}_l\left\{ \right [H_{\langle lb\rangle}^{(2)},\rho_{ab}\rho_l]  \}-i \sum_{l\in nn(a)\neq b}\mbox{Tr}_l\left\{ \right [H_{\langle la\rangle}^{(2)},\rho_{ab}\rho_l]  \},\nonumber \\
&=-i[H^{(1)}_a,\rho_{ab}]-i[H^{(1)}_b, \rho_{ab}]-i[H^{(2)}_{\langle ab\rangle},\rho_{ab}]+\sum_{j=a,b} \mathcal{D}_j(\rho_{ab})+i\frac{V}{4}(q-1)2 m^x [\sigma^x_a,\rho_{ab}]+i\frac{V}{4}(q-1)2 m^x [\sigma^x_b,\rho_{ab}].
\label{app:hierarchy_2_corr}
\end{align}
which is now fundamentally a closed equation for the two-body density matrix describing the nearest neighbor bond $\langle a,b\rangle$. In passing from the second to the third line of Eq.~\eqref{app:hierarchy_2_corr} we note the explicit appearance of the coordination number $q$ in the equation. This describes the interactions in mean-field between the sites $a$ and $b$ with the remaining $q-1$ neighbouring sites, as commented before \eqref{app:hierarchy_2_corr}. The correlations on the single bond $\langle a,b\rangle$ are treated, instead, exactly. In the previous equation, we also exploited translation invariance so that the order parameter $m^x=\Braket{\sigma^{x}_l}/2$ does not depend on the site $l$. From the second to the third line of \eqref{app:hierarchy_2_corr} we also used the property following from the cyclic invariance of the trace
\begin{align}
\mbox{Tr}_l \left\{\left[H^{(2)}_{\langle al \rangle},\rho_{abl}\right] \right\}&=\mbox{Tr}_l \left\{\left[H^{(2)}_{\langle al\rangle},\rho_{ab}\rho_l\right] \right\}=\mbox{Tr}_l \left\{ -\frac{V}{4}\sigma^x_a \sigma^x_l \rho_{ab}\rho_l + \frac{V}{4}\rho_{ab}\rho_l \sigma^x_a \sigma^x_l  \right\}= \nonumber \\
&=-\frac{V}{4}\mbox{Tr}_l\left\{\sigma^x_l \rho_l \right\}[\sigma^x_a, \rho_{ab}]=-\frac{V}{4}2 m^x[\sigma^x_b,\rho_{ab}].
\label{app:commutator_2b_identity}
\end{align}
The most general expansion of the two-body density matrix $\rho_{ab}$ in terms of the Pauli matrices $(\mathbb{I}_{a,b},\sigma^{\mu}_{a,b})$ ( they form a basis of the Hilbert space at the sites $a,b$, with $\mu=x,y,z$) is identified by $15$ independent real numbers (since the density matrix is hermitian and has unit trace). We denote these numbers as $\alpha_{\mu,\nu} \in \mathbb{R}$, with $\mu,\nu=\{I,x,y,z\}$ and we accordingly write the two-body density matrix as
\begin{equation}
\rho_{ab}=\frac{\mathbb{I}_a \mathbb{I}_b}{4}+\frac{1}{4}\left(\sum_{\mu,\nu\in\{x,y,z\}}\alpha_{\mu,I}\sigma^{\mu}_a\,\mathbb{I}_b + \sum_{\mu,\nu\in\{x,y,z\}}\alpha_{I,\nu}\mathbb{I}_a\,\sigma^{\nu}_b +\sum_{\mu,\nu\in\{x,y,z\}}\alpha_{\mu,\nu}\sigma^{\mu}_a\,\sigma^{\nu}_b\right)
\label{eq:two_state_generic}
\end{equation}
Here, the tensor product is always implied between operators defined on the two different lattice sites $a$ and $b$. The physical meaning of the coefficients  $\alpha_{\mu,\nu}$ is readily understood since they are given, as consequence of the orthogonality relation among the Pauli matrices (on the same site) $\mbox{Tr}[\sigma^{\mu}\sigma^{\nu}]=2\delta_{\mu,\nu}$, by
\begin{align}
4m^{\mu,\nu}&=\alpha_{\mu,\nu}=\Braket{\sigma^{\mu}_a \sigma^{\nu}_b}=\mbox{Tr}[\rho^c \sigma^{\mu}_a \sigma^{\nu}_b]=\mbox{Tr}[\rho_{ab} \sigma^{\mu}_a \sigma^{\nu}_b],  \, \, \, \quad \qquad (\mbox{two-point correlation functions}), \label{app:correlation_parameters} \\
2 m^{\mu}&=\alpha_{\mu,I}=\alpha_{I,\mu}=\Braket{\sigma^{\mu}_a}=\Braket{\sigma^{\mu}_b}=\mbox{Tr}[\rho_{ab}\sigma^{\mu}_a]=\mbox{Tr}[\rho_a \sigma^{\mu}_a] \quad (\mbox{one-point functions}).
\label{app:order_pam_parameters}
\end{align}
The parameters $\alpha_{\mu,I}=2 m^{\mu}$ in Eq.~\eqref{app:order_pam_parameters} are equal to the mean magnetization in the direction $\mu=x,y,z$. The factor $2$ in the relation between $\alpha_{\mu,I}$ and $m^{\mu}$ is consistent with the definition of the magnetization adopted in the main text (cf. Eq.~(5) therein). Note that as a consequence of translational invariance one has $\alpha_{\mu,I}=\alpha_{I,\mu}$. The number of independent real numbers parameterizing the reduced two-body state can be therefore reduced to twelve. The three parameters $m^{\mu}$, namely, correspond to the three order parameter components, which are used within the mean-field approximation \eqref{app:mean_field_state}. Differently from this case, here, we take the two-point functions $4m^{\mu,\nu}=\alpha_{\mu,\nu}$ in Eq.~\eqref{app:correlation_parameters} as independent parameters since two-body correlations on the bond $\langle a,b \rangle$ are exactly accounted for by the state \eqref{app:correlations_trial}. The twelve corresponding equation of motions are obtained by inserting the parametrization \eqref{eq:two_state_generic} into \eqref{app:hierarchy_2_corr} and comparing the resulting left and right-hand side component-wise in the decomposition in terms of the Pauli matrices. The resulting equations read 
\begin{subequations}
\begin{align}
\frac{\mbox{d}m^x}{\mbox{d}t}&=-\Omega m^y-\frac{\kappa}{2}m^x, \label{app:corr1} \\
\frac{\mbox{d}m^y}{\mbox{d}t}&=\Omega m^x+V(q-1)m^x m^z+V m^{z,x}-\frac{\kappa}{2}m^y,\label{app:corr2} \\
\frac{\mbox{d}m^z}{\mbox{d}t}&=-V(q-1)m^x m^y -V m^{y,x}-\kappa\left(\frac{1}{2}+m^{z}\right), \label{app:corr3}\\
\frac{\mbox{d}m^{x,x}}{\mbox{d}t}&=-\Omega (m^{y,x}+m^{x,y})-2\kappa m^{x,x},\label{app:corr4}\\
\frac{\mbox{d}m^{x,y}}{\mbox{d}t}&=-\Omega(m^{y,y}-m^{x,x})+V m^x(q-1)m^{z,x}+\frac{V}{4}m^z-2\kappa m^{x,y},\label{app:corr5}\\
\frac{\mbox{d}m^{x,z}}{\mbox{d}t}&=-\Omega m^{y,z}-V m^x(q-1)m^{x,y}-\frac{V}{4}m^y-2\kappa m^{x,z}-\frac{\kappa}{2} m^x,\label{app:corr6} \\
\frac{\mbox{d}m^{y,x}}{\mbox{d}t}&=\Omega (m^{x,x}-m^{y,y}) + V(q-1)m^x m^{z,x}+\frac{V}{4}m^z-2\kappa m^{y,x},\label{app:corr7}\\
\frac{\mbox{d}m^{y,y}}{\mbox{d}t}&=\Omega (m^{x,y}+m^{y,x}) + V m^x(q-1)(m^{y,z}+m^{z,y})-2\kappa m^{y,y}, \label{app:corr8}\\
\frac{\mbox{d}m^{y,z}}{\mbox{d}t}&=\Omega m^{x,z} -V m^x(q-1)(m^{y,y}-m^{z,z})-2\kappa m^{y,z}-\frac{\kappa}{2} m^y,\label{app:corr9}\\
\frac{\mbox{d}m^{z,x}}{\mbox{d}t}&=-\Omega m^{z,y}-V(q-1)m^x m^{y,x}-\frac{V}{4}m^y-2\kappa m^{z,x}-\frac{\kappa}{2} m^x, \label{app:corr10}\\
\frac{\mbox{d}m^{z,y}}{\mbox{d}t}&=\Omega m^{z,x}+V m^x(q-1)(m^{z,z}-m^{y,y})-2\kappa m^{z,y}-\frac{\kappa}{2} m^y, \label{app:corr11}\\
\frac{\mbox{d}m^{z,z}}{\mbox{d}t}&=-V m^x(q-1)(m^{z,y}+m^{y,z})-\kappa(m^z+2 m^{z,z}). \label{app:corr12}
\end{align}
\label{app:correlations_eom_final}%
\end{subequations}
\noindent The equations \eqref{app:correlations_eom_final} have been used to derive the stationary phase diagram reported in Fig.~2(b) of the main text. They represent a $1/q$ ($q=2d$ is the coordination number) correction to the mean-field approximation, which is therefore especially relevant in low spatial dimensions. As a matter of fact, we see that in the three equations \eqref{app:corr1}-\eqref{app:corr3} the correlations of the single bond $\langle a,b\rangle$ are treated exactly and this gives raise to the terms proportional to $\sim V m^{z,x}$ and $\sim V m^{y,x}$ in \eqref{app:corr2} and \eqref{app:corr3}, respectively. The interaction with the remaining $q-1$ neighbouring sites is treated, instead, in a mean-field approximation and it leads to the terms $\sim V(q-1) m^x m^z$ and $V(q-1) m^x m^y$ in \eqref{app:corr2} and \eqref{app:corr3}, respectively. In low dimension, for instance in $d=1$ as discussed in the main text, the two contributions have the same weight and therefore spatial correlations impact significantly on the phase diagram. 

In high space dimensions, $q\gg1$, instead, correlations from the single bond $\langle a,b\rangle$ become less and less significant and the equations \eqref{app:corr4}-\eqref{app:corr12} for the two-point correlation functions can be neglected. The present approach then becomes equivalent to mean-field, which is obtained as a special case of \eqref{app:correlations_eom_final} for $q\gg 1$ by factorizing the two-point functions as $m^{\mu,\nu}=m^{\mu} m^{\nu}$. The one-point functions $m^{\mu}=2 \alpha_{\mu,I}$ are the sole independent parameters and the corresponding equations \eqref{app:corr1}-\eqref{app:corr3} become equivalent to those obtained within mean-field: 
\begin{subequations}
\begin{align}
\frac{\mbox{d}m^x}{\mbox{d}t}&=-\Omega m^y-\frac{\kappa}{2}m^x, \\
\frac{\mbox{d}m^y}{\mbox{d}t}&=\Omega m^x+´V q\, m^x m^z-\frac{\kappa}{2}m^y,\\
\frac{\mbox{d}m^z}{\mbox{d}t}&=-V q\, m^x m^y -\kappa\left(\frac{1}{2}+m^{z}\right),    
\end{align}
\label{app:mean_field_eom_final}%
\end{subequations}
These equations coincide with Eq.~(6) of the main text, where the one-dimensional case is considered ($q=2$). The corresponding numerical solution in one dimension ($q=2$) leads to the mean-field stationary phase diagram in Fig.~2(a) of the main text.

\section{Quantum cellular automaton simulations}
In this section we provide additional information on the numerical simulations of our quantum cellular automaton (QCA) dynamics and training presented in Figs.~3 and 4 of the main text. 

\subsection{Initial State Sampling}
We here explain how the sampling of the initial state $\rho_0$ adopted in the simulations of the main text. The mean-field and nearest neighbor correlation analysis [see Fig.~2 of the main text] shows that the magnetization order parameter  $m_t^x=\Tr(\sum_{k=1}^N \sigma_k^x \rho_t)/(2N) \in [-0.5,0.5]$  is a relevant quantity to study ergodicity breaking in our Ising model inspired QCA. A \textit{uniform} sampling of the initial states, with respect to their value for the order parameter $m_0^x$, is thus fundamental in order not to introduce any bias in the magnetization histograms. This allows to provide a faithful assessment of the presence of transient ergodicity breaking and classification of input states in QCA [see Figs.~3 and 4 of the main text]. Concretely, we choose $n=1000$ initial product states $\rho_0^l = \bigotimes_{k=1}^N | \psi^l \rangle \langle \psi^l |, \ l \in \{ 1,\ldots,n \}$ with 
\begin{equation}
    | \psi^l \rangle = \cos(\frac{\theta^l}{2}) |0 \rangle + e^{i \phi^l} \sin(\frac{\theta^l}{2}) |1 \rangle .
\end{equation}
The angles are parametrized in terms of the magnetizations $m_{0,l}^x,m_{0,l}^y,m_{0,l}^z$ via 
\begin{equation}\label{eq:mb1}
    \theta^l = \arccos(2m_{0,l}^z) \ , \quad \phi^l = \mathrm{arctan2}(m_{0,l}^y,m_{0,l}^x),  \quad  \mathrm{and} \quad \left(m_{0,l}^x \right)^2 + \left(m_{0,l}^y \right)^2 + \left(m_{0,l}^z \right)^2 = 1/4.
\end{equation}
The second equation employs the $\mbox{arctan2}$ function, which is defined as
\begin{equation}
\mathrm{arctan2}(y,x) =
\begin{cases}
\arctan\!\left(\tfrac{y}{x}\right), & x > 0, \\[1em]
\arctan\!\left(\tfrac{y}{x}\right) + \pi, & x < 0,\; y > 0, \\[1em]
\pm \pi, & x < 0,\; y = 0, \\[1em]
\arctan\!\left(\tfrac{y}{x}\right) - \pi, & x < 0,\; y < 0, \\[1em]
+\tfrac{\pi}{2}, & x = 0,\; y > 0, \\[0.5em]
-\tfrac{\pi}{2}, & x = 0,\; y < 0.
\end{cases}
\end{equation}
This function has to be used as the underlying expression for $\phi^l \in (-\pi,\pi]$ is $\tan(\phi^l)=m_{0,l}^y / m_{0,l}^x$. The inverse of $\tan$ does not distinguish the cases with, e.g., $\mathrm{sign}(m_{0,l}^y)=1, \mathrm{sign}(m_{0,l}^x)=1$ and $\mathrm{sign}(m_{0,l}^y)=-1, \mathrm{sign}(m_{0,l}^x)=-1$. The $\mathrm{arctan2}$ accounts for this and gives the correct quadrant for the angle $\phi^l$.
The last equation in Eqs.~\eqref{eq:mb1} follows from the fact that single-qubit pure states can be parametrized   
\begin{equation}
    | \psi^l \rangle \langle \psi^l | = \frac{\mathbb{I} + r_1 \sigma^x + r_2 \sigma^y + r_3 \sigma^z }{2},
\end{equation}
with $\sqrt{r_1^2+r_2^2+r_3^2}=1$ (they lie on the surface of the Bloch sphere). Here $r_1=2m_{0,l}^x,r_2=2m_{0,l}^y,r_3=2m_{0,l}^z$. We choose product states as they can be easily implemented as matrix product states. Given this parametrization of the $n=1000$ states, we fix them by uniformly sampling $m_{0,l}^x$ in the interval $[0,0.5]$. The other magnetization components then follow by computing 
\begin{equation}
    R= \sqrt{1/4 - \left( m_l^x \right)^2},
\end{equation}
and 
\begin{equation}
    m_l^y = R \sin(\alpha) \ , \quad m_l^z = R \cos(\alpha),
\end{equation}
with $\alpha \in [0, 2 \pi]$ randomly. To get the other $1000$ states $\tilde{\rho}_0^l$, which uniformly cover the interval $m^x \in [-0.5,0]$, we apply a $\mathbb{Z}_2$ unitary transformation $\mathcal{P}[\bullet]=\prod_{k=1}^N \sigma_k^z \bullet \prod_{k=1}^N \sigma_k^z$ to $\rho_0^l$: 
\begin{equation}
\tilde{\rho}_0^l = \mathcal{P}[\rho_0^l].
\label{eq:z2_supp}
\end{equation}
This ensures that the initial magnetization $\tilde{m}_{0,l}^{x}$ associated to the states $\tilde{\rho}_0^l$
\begin{equation}
    \tilde{m}_{0,l}^{x} =\Tr(\frac{1}{2N}\sum_{k=1}^N \sigma_k^x \tilde{\rho}_0^l) =\Tr(\frac{1}{2N}\sum_{k=1}^N \mathcal{P}[\sigma_k^x] \rho_0^l) = - m_{0,l}^{x},
\end{equation}
is opposite to the magnetization $m_{0,l}^{x}$ of the states $\rho_0^l$. Moreover, one can as well show that the QCA we consider in this work (Eqs.~(1) and (2) of the main text) are weakly $\mathbb{Z}_2$-symmetric under $\mathcal{P}$. This means that upon writing  as $\rho_t= \Lambda [\rho_{t-1}]$, with the linear map $\Lambda[\bullet]$ defined by the recurrence relation (1) of the main text, one has
\begin{equation}
    \Lambda [\mathcal{P}[\rho_{t-1}]] =\mathcal{P}[\Lambda [\rho_{t-1}]]=\mathcal{P}[\rho_{t}].
\label{sm:weak_z2}
\end{equation}
As a consequence, the magnetization $m_{t,l}^{x}$ on layer $t$ obtained from the initial state $\rho_0^l$
\begin{equation}
    \tilde{m}_{t,l}^{x} =\Tr(\frac{1}{2N}\sum_{k=1}^N \sigma_k^x \tilde{\rho}_t^l) =\Tr(\frac{1}{2N}\sum_{k=1}^N \sigma_k^x \Lambda[\mathcal{P}[\rho_{t-1}^l]]) = \Tr(\frac{1}{2N}\sum_{k=1}^N \mathcal{P}[\sigma_k^x]  \rho_{t}^l) = -m_{t,l}^{x} ,
\label{sm:symmetry_layer}
\end{equation}
is the opposite of the one $\tilde{m}_{t,l}^{x}$ on the same layer $t$, but obtained from the initial state $\tilde{\rho}_0^l$. In the previous equation \eqref{sm:symmetry_layer}, we used that $\tilde{\rho}_t^l$ is obtained by applying $t$ times the map $\Lambda$ on the state $\tilde{\rho}_0^l$. The weak symmetry relation \eqref{sm:weak_z2} is then used $t$ times together with cyclic invariance of the trace in order to bring the action of $\mathcal{P}$ onto the magnetization operator. The time evolution of magnetization curves is thus symmetric with respect to the time layer axis. As a last step, necessary for the tensor-network simulations, the states $\rho_0^l,\tilde{\rho}_0^l$ are then vectorized and written as matrix product states [see the section here below for details on the tensor network implementations of QCAs].

\subsection{Stability of tensor network simulations with respect to changes in bond dimension}
Our QCA propagates initial states via the recurrence relation (1) in the main text. In our classical simulations, we implement this using matrix product state based techniques. We give here a brief summary of the method, while a detailed discussion can be found in Refs.~\cite{orus2014,paeckel2019,gillman2021b,boneberg2025}; we essentially discard irrelevant information at the level of the coefficients of states and operators. How much information is kept in the matrix product representation of states and operators is quantified by the so-called bond-dimension $\chi$. At the same time, this quantity provides a measure for the computational costs. The higher the bond dimension, the better the accuracy, but the higher the computational costs. Here, we want to demonstrate that, although the bond dimensions that we use in this work are rather small, they still capture the essential physics. Transient ergodicity breaking is therefore a genuine feature of the QCA dynamics and it is not a numerical artifact due to the choice of a too small bond dimension.
\begin{figure}[H]
    \centering
    \includegraphics[width=\linewidth]{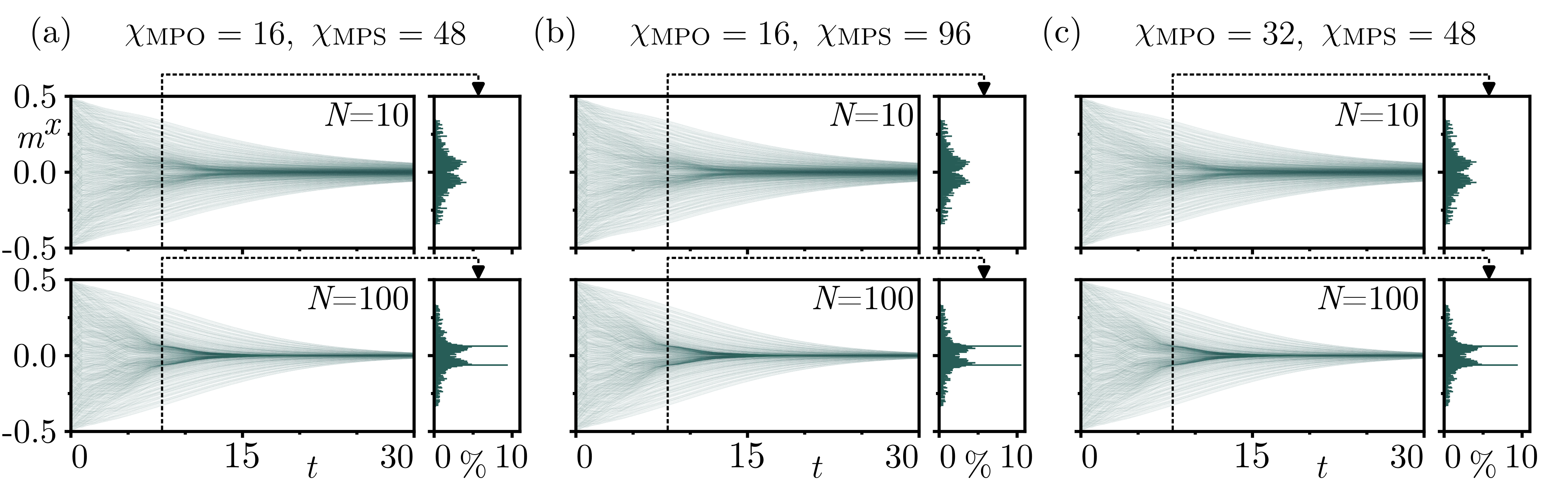} 
    \caption{\textbf{Stability of tensor-network simulations of QCA dynamics with respect to changes of the bond dimension.} The upper panels show the QCA evolution for $N=10$ and the histogram of magnetizations $m_8^x$ at time layer $t=8$. The lower panels show the same for $N=100$. (a) This panel is equal to figure 3 of the main text and it is reported for the sake of comparison. (b) Analogous to panel (a), we plot the QCA simulation results for an increased matrix product state bond dimension $\chi_{\mathrm{MPS}}=96$. There are no visible changes in the evolution. Only the histogram for $N=100$ shows slightly higher peaks. (c) Analogous to (a) the QCA simulations for increased matrix product operator bond dimension $\chi_{\mathrm{MPO}}=32$. There are no visible changes compared to the results in Fig.~4 of the main text.}
    \label{supfig:1}
\end{figure}

For the simulations in Figs.~3 and 4 of the main text we chose a matrix product state bond dimension $\chi_{\mathrm{MPS}}=48$ and a matrix product operator bond dimension $\chi_{\mathrm{MPO}}=16$. In Fig.~\ref{supfig:1}, we illustrate that the QCA evolutions shown in Fig.~3 of the main text do not qualtitatively change when we increase the matrix product state bond dimension to $\chi_{\mathrm{MPS}}=96$ and the matrix product operator bond-dimension to $\chi_{\mathrm{MPO}}=32$. The quantitative corrections are only minor. In fact, throughout our simulations, the square root of the total sum of the squared discarded singular values (which quantifies the discarded information) for $\chi_{\mathrm{MPO}}=16$ is of the order of machine precision and we therefore regard such a bond dimension for the matrix product operators as sufficient. 

\begin{figure}[H]
    \centering
    \includegraphics[width=\linewidth]{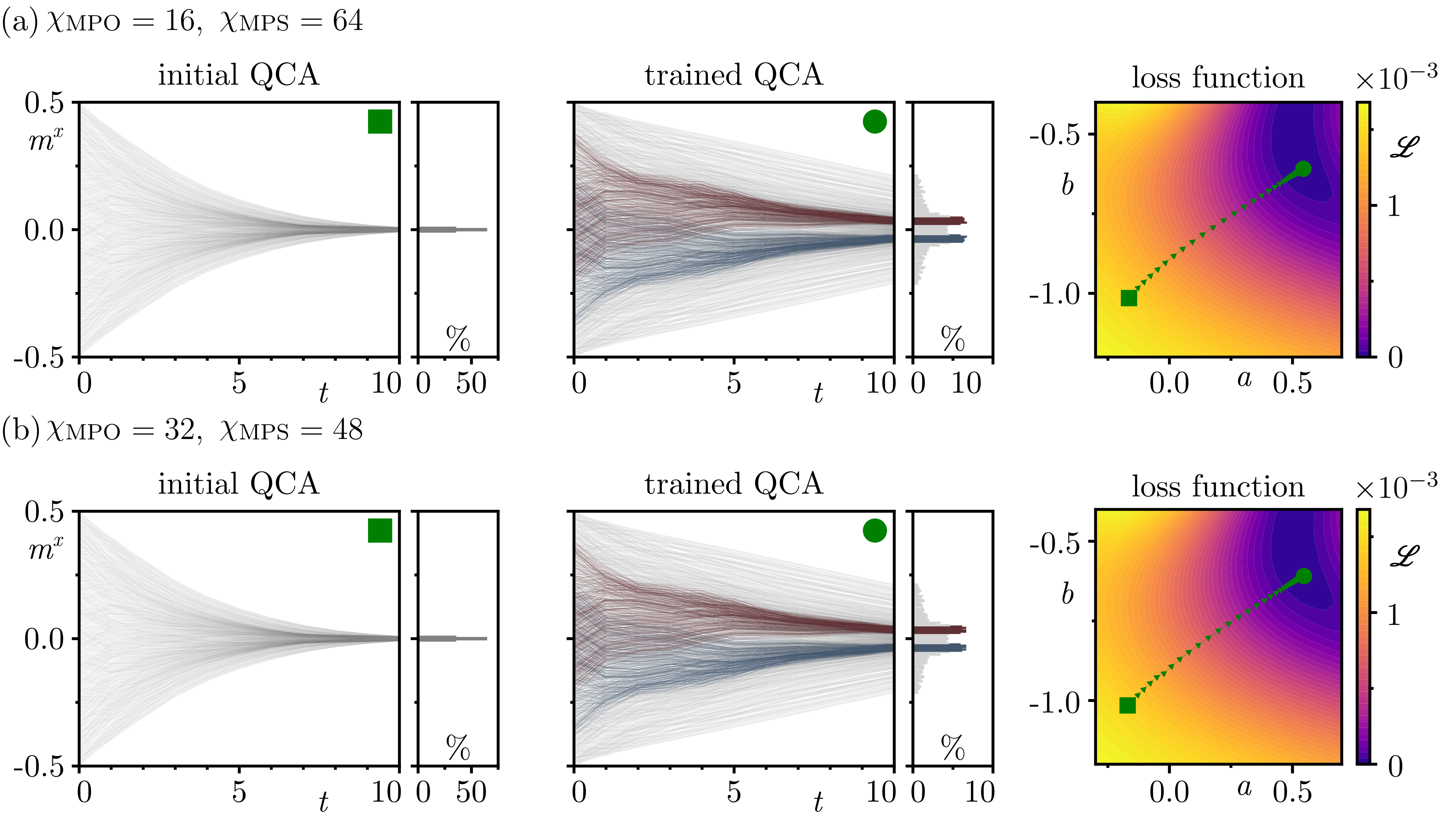} 
    \caption{\textbf{Stability of training results with respect to changes in the bond dimension.} (a) Training results for increased matrix product state bond dimension $\chi_{\mathrm{MPS}}=64$ and $\chi_{\mathrm{MPO}}=16$. As can be seen the evolution of the initial and trained QCA, the loss landscape, as well as the line of gradient descent are in close agreement with the original simulations for $\chi_{\mathrm{MPS}}=48$ and $\chi_{\mathrm{MPO}}=16$ in Fig.~4 of the main text but. (b) Training results for increased matrix product operator bond dimension $\chi_{\mathrm{MPO}}=32$ and $\chi_{\mathrm{MPS}}=48$. Also here the results do not deviate qualitatively from the ones in Fig.~4 of the main text. In these plots we trained with 50 repetitions of the update and a learning rate $\epsilon = 80$. As well as in the main text, we fixed $N=20$.}
    \label{supfig:2}
\end{figure}

Moreover, in Fig.~\ref{supfig:2} we show the results of our training procedure if we increase the matrix product state bond dimension to $\chi_{\mathrm{MPS}}=64$ [panel (a)] and the matrix product operator bond dimension to $\chi_{\mathrm{MPO}}=32$ [panel (b)]. As can be seen, the training results are qualitatively identical to those in Fig.~4 of the main text. It is important to emphasize that we simulate the initial QCA, the trained QCA, as well as the training process and the loss function (defined in Eq.~(8) of the main text) in the whole $a-b$ plane with the reported higher values for the bond dimensions.

\end{document}